\documentclass[11pt]{article}
\usepackage{amssymb}
\usepackage{amsmath}
\usepackage{amstext}
\usepackage{graphicx,epsfig}
\usepackage{epsfig}
\usepackage{verbatim} 
\usepackage{fancybox}
\usepackage{color}
\usepackage{ulem}
\usepackage{enumitem}
\usepackage{subfigure}
\usepackage{bbm}
\usepackage{parskip}
\usepackage{dsfont}
\usepackage[numbers,sort&compress]{natbib}

\newcommand{\Comment}[1]{{}}
\definecolor{darkblue}{rgb}{0.15,0.35,0.55}
\definecolor{reddish}{rgb}{0.65, 0.2, 0.2}
\usepackage[linktocpage=true]{hyperref}
\hypersetup{
colorlinks=true,
citecolor=darkblue,
linkcolor=reddish,
urlcolor=darkblue,
pdfauthor={},
pdftitle={},
pdfsubject={}
}

\newcommand{\nn}{\nonumber \\ }

\def\({\left(}
\def\){\right)}

\newcommand{\be}{\begin{equation}}
\newcommand{\ee}{\end{equation}}
\newcommand{\bea}{\begin{align}}
\newcommand{\eea}{\end{align}}

\newcommand{\rd}{{\rm d}}

\newcommand{\Tr}{\, {\rm Tr} \, }

\def\gsim{ \lower .75ex \hbox{$\sim$} \llap{\raise .27ex \hbox{$>$}} }
\def\lsim{ \lower .75ex \hbox{$\sim$} \llap{\raise .27ex \hbox{$<$}} }

\def\xyma{\xymatrix@M.7em}
\def\xymas{\xymatrix@M.1em}
\newcommand{\ba}{\begin{eqnarray}}
\newcommand{\ea}{\end{eqnarray}}

\usepackage{tikz}
\usetikzlibrary{decorations}
\pgfdeclaredecoration{complete sines}{initial}
{
    \state{initial}[
        width=+0pt,
        next state=upsine,
        persistent precomputation={\pgfmathsetmacro\matchinglength{
            \pgfdecoratedinputsegmentlength / int(\pgfdecoratedinputsegmentlength/\pgfdecorationsegmentlength)}
            \setlength{\pgfdecorationsegmentlength}{\matchinglength pt}
        }] {}
    \state{upsine}[width=\pgfdecorationsegmentlength,next state=downsine]{
        \pgfpathsine{\pgfpoint{0.25\pgfdecorationsegmentlength}{0.5\pgfdecorationsegmentamplitude}}
        \pgfpathcosine{\pgfpoint{0.25\pgfdecorationsegmentlength}{-0.5\pgfdecorationsegmentamplitude}}
    }
    \state{downsine}[width=\pgfdecorationsegmentlength,next state=upsine]{
        \pgfpathsine{\pgfpoint{0.25\pgfdecorationsegmentlength}{-0.5\pgfdecorationsegmentamplitude}}
        \pgfpathcosine{\pgfpoint{0.25\pgfdecorationsegmentlength}{0.5\pgfdecorationsegmentamplitude}}
}
    \state{final}{}
}

\title{}
\author{}

\numberwithin{equation}{section}

\begin{document}
%
\renewcommand{\thefootnote}{\fnsymbol{footnote}}
~
\vspace{2.5truecm}
\begin{center}
{\LARGE \bf{Aspects of Galileon Non-Renormalization}}
\end{center} 
 \vspace{1truecm}
\thispagestyle{empty} \centerline{
{\Large  {Garrett Goon,${}^{\rm a}$}}
{\Large  {Kurt Hinterbichler,${}^{\rm b}$}}
{\Large {Austin Joyce,${}^{\rm c}$}}
{\Large and} 
{\Large Mark Trodden${}^{\rm d}$}
                                                           }

\vspace{.5cm}

\centerline{{\it ${}^{\rm a}$Department of Applied Mathematics and Theoretical Physics,}}
\centerline{{\it Cambridge University, Cambridge, CB3 0WA, UK}}

\vspace{.25cm}
\centerline{{\it ${}^{\rm b}$Perimeter Institute for Theoretical Physics,}}
 \centerline{{\it 31 Caroline St. N, Waterloo, Ontario, Canada, N2L 2Y5}}

\vspace{.25cm}
\centerline{\it ${}^{\rm c}$Enrico Fermi Institute and Kavli Institute for Cosmological Physics,}
\centerline{\it University of Chicago, Chicago, IL 60637}

 \vspace{.25cm}
 \centerline{{\it ${}^{\rm d}$Center for Particle Cosmology, Department of Physics and Astronomy,}}
 \centerline{{\it University of Pennsylvania, Philadelphia, PA 19104, USA}}

 \vspace{.5cm}
\begin{abstract}
\noindent
We discuss non-renormalization theorems applying to galileon field theories and their generalizations. Galileon theories are similar in many respects to other derivatively coupled effective field theories, including general relativity and $P(X)$ theories. In particular, these other theories also enjoy versions of non-renormalization theorems that protect certain operators against corrections from self-loops. However, we argue that the galileons are distinguished by the fact that they are not renormalized even by loops of other heavy fields whose couplings respect the galileon symmetry.
\end{abstract}

\newpage

\setcounter{tocdepth}{2}
\tableofcontents
\newpage
\renewcommand*{\thefootnote}{\arabic{footnote}}
\setcounter{footnote}{0}

\section{Introduction}
The galileons \cite{Nicolis:2008in} are a fascinating class of higher-derivative scalar effective field theories which display rich and varied structure and phenomenology. They have elegant geometrical origins as the description of brane fluctuations in the DGP model \cite{Dvali:2000hr,Luty:2003vm} (further elaborated upon in \cite{deRham:2010eu,Goon:2011qf,Goon:2011uw,Goon:2012dy}), describe the helicity zero mode of a ghost-free interacting massive spin-2 field \cite{deRham:2010ik,deRham:2010kj}, are the key players in interesting IR modifications of GR which display Vainshtein screening \cite{Vainshtein:1972sx,Babichev:2013usa} near heavy objects~\cite{Nicolis:2004qq,Nicolis:2008in}, and possess an S-matrix with many special properties \cite{Kampf:2014rka,Cheung:2014dqa,Cachazo:2014xea,Cheung:2015ota,Hinterbichler:2015pqa}.\footnote{The galileons also possess unusual features: for solutions around heavy sources, perturbations can propagate superluminally \cite{Nicolis:2008in,Goon:2010xh} (though this can be alleviated in other examples~\cite{Berezhiani:2013dw,Gabadadze:2014gba}), and, treated in isolation, there are arguments that galileons have no local, Lorentz invariant UV completion \cite{Adams:2006sv,Bellazzini:2016xrt} (however, when incorporated into full massive gravity, these obstructions are lifted in some cases \cite{Cheung:2016yqr}).  See \cite{Dvali:2010jz,Dvali:2010ns,Codello:2012dx,Brouzakis:2013lla,Cooper:2013ffa,Keltner:2015xda} for interpretations of these features in terms of non-standard UV completions.}

 In this paper we focus on another property of galileons: their non-renormalization theorem. Certain galileon operators are not renormalized by galileon loops~\cite{Luty:2003vm, Hinterbichler:2010xn}.  We are interested in understanding both the importance of this fact and how it compares to other, superficially similar, non-renormalization theorems obeyed by other effective field theories.
 
The simplest example of a galileon is a single scalar field, $\phi(x)$, which obeys a shift symmetry linear in coordinates,
\be
 \phi(x)\longmapsto \phi(x)+c+b_{\mu}x^{\mu}\ , \label{GalileonSymmetryIntro}
\ee
 with $c,b_{\mu}$ constant.  Any term built out of $\partial_{\mu}\partial_{\nu}\phi$, and its derivatives, will be strictly invariant under \eqref{GalileonSymmetryIntro}.  However, there also exist special operators with {fewer} than two derivatives per $\phi$, which are not strictly invariant, but rather are invariant up to a total derivative.  The cubic galileon interaction is the canonical term of this type
\be
 S_{\rm cubic}=\int\rd^{4}x\left(-\frac{1}{2}(\partial\phi)^{2}-\frac{1}{\Lambda^{3}}(\partial\phi)^{2}\square\phi\right ),\label{CubicGalileonIntro}
\ee
 with $\Lambda$ some strong coupling scale.
 These special operators are reviewed in Sec.$\!$ \ref{Sec:NonRenormalizationTheorem}.

The statement of the non-renormalization theorem is that loops of galileon fields only serve to renormalize the higher derivative operators built from $\partial_{\mu}\partial_{\nu}\phi$. For example, $\Lambda$ in \eqref{CubicGalileonIntro} doesn't run.  This is in accord with general folklore stating that terms invariant only up to a total derivative are typically protected in some way against quantum corrections.   Examples are the Wess--Zumino--Witten (WZW) term \cite{Witten:1983tw} in the chiral Lagrangian and Chern--Simons terms in three dimensional Yang--Mills theory \cite{Deser:1981wh} whose coefficients don't run and, further, are quantized \cite{Witten:1983tw,Witten:1988hf}.  The special galileon operators to which the non-renormalization theorem applies are, in fact, an analogue of the WZW term \cite{Goon:2012dy}, albeit there is no argument to suggest their coefficients should be quantized.  
 
One puzzle is that the theorem is simultaneously non-trivial and trivial, in some sense.  It is non-trivial in that there exists a diagrammatic proof of the theorem \cite{Hinterbichler:2010xn} which heavily relies on the detailed structure of the special galileon operators.  It is trivial in that the same conclusions also essentially follow from dimensional analysis arguments applied to self-loop graphs in dimensional regularization, with no reference to the detailed form of the galileon operators.  These dimensional analysis arguments can be made for many other massless, derivatively-coupled theories including General Relativity (GR),  $P(X)$ theories and the conformal dilaton field (alternatively known as the conformal galileon \cite{Nicolis:2008in,Creminelli:2010ba}). Certain low-dimension operators in these other theories are also not renormalized by self loops.
 
 What, then, is the non-trivial content of the galileon renormalization theorem?  Are the renormalization properties of galileons qualitatively different from those of GR, $P(X)$ and conformal dilaton, or do they follow similarly from the derivative expansion of effective field theory?
 
In this paper, we argue that the essential difference comes once we consider loops of \textit{heavy} fields which couple to the galileon, $\phi(x)$.  The galileon renormalization theorem implies that the special galileon operators are not renormalized even by heavy fields---provided that they couple in a way which respects the galileon symmetry.  This effect is not visible by considering only self-loops in dimensional regularization, which captures only logarithmic corrections.  It is, loosely speaking, captured by power divergences in graphs.  The detailed diagrammatic proof of the galileon renormalization theorem tells us that the entire quantum contribution to the galileon vanishes, including power corrections.  This suggests that coupling heavy fields to the galileon should not renormalize the galileon operators, and indeed this is what we will find in explicit examples.
 In contrast, coupling heavy fields to GR, $P(X)$ and the conformal dilaton \textit{do} affect the operators in these theories which are not renormalized by the logarithmic part of self loops. In this precise sense the galileon non-renormalization theorem is stronger.
 
 In Sec.$\!$ \ref{Sec:GalileonReview} we review galileon theories in more detail and discuss the detailed version of the non-renormalization theorem.  In Sec.$\!$ \ref{Sec:DimensionalArguments} we review how the non-renormalization theorem follows from dimensional analysis, apply the same arguments to other theories, and discuss the motivations behind our above statements.  In Sec.$\!$ \ref{Sec:CouplingHeavyFields} we illustrate how heavy physics affects the operators in these theories by coupling in a massive scalar field and integrating it out. The methods used for integrating out the heavy field are summarized in Appendix \ref{Appendix:RulesForMatrixElements}, and we conclude in Sec.$\!$ \ref{Sec:Conclusion}.
 
\vspace{-.5cm}
\paragraph{Conventions:} Throughout we use mostly plus metric signature. We denote the flat-space d'Alembert operator by $\partial^2\equiv \eta^{\mu\nu}\partial_\mu\phi\partial_\nu\phi$.

\section{Review of Galileons and their Non-Renormalization\label{Sec:GalileonReview}}

In this section we briefly review some of the basic properties of the galileon and the non-renormalization theorem.  A galileon scalar field, $\phi(x)$, has an action which is invariant under the extended shift symmetry \eqref{GalileonSymmetryIntro}.  
In order for the action to be invariant under this symmetry, the interaction terms must involve derivatives. 
Many of the operators invariant under \eqref{GalileonSymmetryIntro}, powers of $\partial^2\phi$ for example, will lead to higher order equations of motion (EOM) and hence will generically run afoul of Ostrogradski's theorem \cite{Ostrogradsky:1850}, leading to instabilities (see \cite{Woodard:2006nt,Woodard:2015zca} for nice reviews). These instabilities are not problematic as long as the theory is treated as an {effective} field theory (EFT) \cite{Burgess:2014lwa,Simon:1990ic,Jaen:1986iz}.

Interestingly, not all operators invariant under the galileon symmetry are of this type; there exist a finite number of operators which have fewer than two derivatives per field and thus are not constructed from the invariant building block $\partial_\mu\partial_\nu\phi$.   In addition, they yield strictly second order equations of motion.  The existence of such operators opens up a regime in which we can reliably study the non-linear, classical phenomena dictated by these terms, while consistently ignoring the effects of the higher derivative operators discussed in the preceding paragraph \cite{Nicolis:2004qq,Endlich:2010zj}. This can be thought of in analogy with Einstein gravity: there is a regime where classical non-linearities are important, for example in the vicinity of the event horizon of a black hole, while quantum mechanical phenomena are (at least from the point of view of local effective field theory) expected to be unimportant.

 In $d$ spacetime dimensions, there are $d+1$ of these special operators, all of which change by a total derivative under \eqref{GalileonSymmetryIntro}. In four dimensions, they take the form \cite{Nicolis:2008in}
\begin{align}  
{\cal L}_1&=\phi \nn
{\cal L}_2&= (\partial\phi)^2 \nn
{\cal L}_3&= \partial^2 {\phi}(\partial {\phi})^2 \nn
{\cal L}_4&= (\partial\phi)^2\left((\partial^2\phi)^2-(\partial_\mu\partial_\nu\phi)^2\right) \nn
{\cal L}_5&= (\partial\phi)^2\left((\partial^2\phi)^3+2(\partial_\mu\partial_\nu\phi)^3-3\partial^2\phi(\partial_\mu\partial_\nu\phi)^2\right)\ .\label{GalileonTerms}
\end{align}
These operators can be compactly written using the Levi--Civita symbol, which makes many of their properties manifest,
\be
{\cal L}_n \propto \epsilon_{\mu_1\cdots\mu_{n-1} \alpha_n\cdots\alpha_4}\epsilon^{\nu_1\cdots\nu_{n-1}\alpha_n\cdots\alpha_4}\phi\partial^{\mu_1}\partial_{\nu_1}\phi\cdots\partial^{\mu_{n-1}}\partial_{\nu_{n-1}}\phi\,.
\label{GalileonTermsWithEpsilons}
\ee
The anti-symmetric structure of the epsilons guarantees that having two derivatives with either a $\mu$ or $\nu$ index acting on a $\phi$ vanishes, making it easy to see that the galileon terms have second order equations of motion and shift under the symmetry~\eqref{GalileonSymmetryIntro} by a total derivative. 

For the remainder of the paper, we will follow common conventions and refer to the special terms in \eqref{GalileonTerms} alone as ``galileons."  All other terms compatible with the galileon symmetry will simply be called ``higher order operators."

\subsection{The Non-Renormalization Theorem\label{Sec:NonRenormalizationTheorem}}

Loops of $\phi$ fields don't renormalize the galileon interactions~\eqref{GalileonTermsWithEpsilons} at any order in perturbation theory.  This was first noted for the cubic galileon theory in \cite{Luty:2003vm} and then extended to the fully general case in \cite{Hinterbichler:2010xn}.
The argument of \cite{Hinterbichler:2010xn} is phrased in terms of the 1PI action and is diagrammatic. We present a simple path integral version of the same argument here.  The detailed form of the galileon interactions is crucial to each of these arguments.

Consider calculating the 1PI effective action via the background field method \cite{Abbott:1981ke} for a galileon Lagrangian, $\mathcal{L}(\phi)=\sum_{i=1}^{5}c_{i}\mathcal{L}_{i}$ with $\mathcal{L}_{i}$ as in \eqref{GalileonTerms} (higher order operators with more derivatives can also be added to the action without altering the conclusions of the following argument).  The effective action, $\Gamma[\bar\phi]$, for an arbitrary field profile $\bar{\phi}(x)$ can be derived by taking the bare Lagrangian $\mathcal{L}(\phi)$, expanding the field about the background $\phi=\bar{\phi}+\varphi$ and path integrating over the fluctuation $\varphi$ keeping {only} bubble diagrams which are 1PI with respect to fluctuation lines:
\begin{align}
\exp i\Gamma[\bar{\phi}]&=\int_{\rm 1PI}\mathcal{D}\varphi\, \exp iS[\bar{\phi}+\varphi]\label{BackgroundFieldPathIntegral}\ .
\end{align}

We would like to verify that there are no corrections to the coefficients of the galileon operators~\eqref{GalileonTerms}.
We perform the replacement $\phi=\bar{\phi}+\varphi$ in the galileon operators \eqref{GalileonTerms}.  When written in terms of $\epsilon$ tensors, it is immediately clear that we can always integrate the result by parts so that every $\bar{\phi}$ factor has exactly two derivatives acting upon it.  For instance, making the replacement in the cubic operator we find that $\mathcal{L}_{3}$ generates terms of the form
\begin{align}
\mathcal{L}_{3}(\bar{\phi}+\varphi)&\supset\ \  \sim \epsilon_{\nu_{1}\nu_{2}\alpha_{3}\alpha_{4}}\epsilon^{\mu _{1}\mu_{2}\alpha_{3}\alpha_{4}}\left (\partial_{\mu_1}\bar\phi\partial^{\nu_{1}}\varphi\partial_{\mu _{2}}\partial^{\nu_{2}}\varphi\right ),
\end{align}
we can then integrate the $\partial^{\nu_{1}}$ derivative by parts, to turn this operator into
\begin{align}
\mathcal{L}_{3}(\bar{\phi}+\varphi)&\supset\ \  \sim \epsilon_{\nu_{1}\nu_{2}\alpha_{3}\alpha_{4}}\epsilon^{\mu _{1}\mu_{2}\alpha_{3}\alpha_{4}}\left (\varphi\partial^{\nu_{1}}\partial_{\mu_1}\bar\phi\partial_{\mu _{2}}\partial^{\nu_{2}}\varphi\right )\ .
\end{align}
It is clear that this argument will generalize to all of the galileon operators; note that this relies crucially on the particular structure of these terms.

In the resulting path integral \eqref{BackgroundFieldPathIntegral}, $S[\bar{\phi}+\varphi]$ thus contains terms involving $\varphi$ with either zero, one or two derivatives acting upon it, but it depends on $\bar{\phi}$ \textit{strictly} through the combination $\partial^{2}\bar\phi$.  Path integrating over $\varphi$, the result $\Gamma[\bar\phi]$ will be build from propagators and vertices which depend on $\partial_\mu\partial_\nu\bar\phi$ and derivatives thereof.  All generated terms must therefore have at least two derivatives per $\bar{\phi}$, while the galileons \eqref{GalileonTerms} have fewer than this, hence the galileons are not renormalized by self loops.\footnote{Because the galileon is massless, $\Gamma$ is expected to have non-local terms involving objects like $\log\partial^2$ and $\partial^{-2}$, so one might worry that powers of inverse $\partial^2$'s could somehow ``cancel out" the derivatives acting on $\bar{\phi}$'s.  The dimensional analysis arguments of the next section ensure that this doesn't happen. 
}  Similar arguments hold in scalar-tensor generalizations of the galileon \cite{deRham:2012ew}, and underlie the technical naturalness of ghost-free massive gravity \cite{deRham:2013qqa}.

\section{Non-Renormalization and Power Counting\label{Sec:DimensionalArguments}}

We now argue that the non-renormalization theorem as previously stated also follows as a simple statement about power counting in effective field theory, and that essentially all derivatively-coupled theories enjoy a similar non-renormalization for their leading operators.

\subsection{General Power Counting}

To make invariant statements about non-renormalization in EFTs, we will want to be able to estimate the way that various diagrams scale with the external momenta of particles. These estimates will allow us to quickly check whether an operator can be renormalized by a loop diagram.

We will want to make estimates of the scaling of observables in theories of the form
\be
\mathcal L_{\rm eff} = \Lambda^4\sum_j\frac{c_j}{\Lambda^{f_j+d_j}}{\cal O}_j\left(\phi,\partial \right)
\label{GenericLagrangian}.
\ee
Here ${\cal O}_j\left(\phi,\partial\right)$ stands for any operator built out of $\phi$ and derivatives thereof, $c_j$ are order one dimensionless coefficients, and $f_j$ and $d_j$ count the number of fields and derivatives appearing in $\mathcal{O}_{j}$, respectively. In \eqref{GenericLagrangian}, we have assumed that only one scale $\Lambda$ enters the Lagrangian, for simplicity.  The extension to multiple scales is straightforward, but unnecessary for our interests.  Throughout the body of this paper, we will only discuss massless theories (apart from a short discussion in the conclusions of how the following estimates and results change when the theory is massive).

In the theory \eqref{GenericLagrangian}, the momentum dependence of an $N$-point scattering amplitude\footnote{The special case of $N=2$ can be thought of as the amputated vacuum polarization diagram.} (or off-shell amputated correlator) $\mathcal{M}^{(N)}$, can easily be estimated, following \cite{Burgess:2007pt}. 
The overall mass dimension of the amplitude is $4-N$, every loop integral leads to an integration $\sim \int\rd^{4}k$, and every internal line contributes $\sim \frac{1}{k^{2}}$ (we only consider bosonic theories).  The only other way factors of momenta can appear is through derivatives in the interaction terms, $\mathcal{O}_{j}\left (\phi,\partial\right )$. Denoting the number of loops in the diagram by $L$,  the number of internal lines by $I$, and the number of vertices with $i$ lines and $n$ derivatives by $V_{(i,n)}$, we find that the amplitude scales as $\sim k^{4L-2I+\sum_{i,n}nV_{(i,n)}}$. Dividing by powers of $\Lambda$ to ensure the correct dimension, we obtain the estimate
\begin{align}
\mathcal{M}^{(N)}\sim \Lambda^{4-N}\left (\frac{k}{\Lambda}\right )^{4L-2I+\sum_{i,n}nV_{(i,n)}}\label{FirstAmplitudeEstimate} \ ,
\end{align}
where $k$ represents some combination of the external momenta.
The result \eqref{FirstAmplitudeEstimate} can simplified somewhat with the use of simple graph-theoretic identities~\cite{Burgess:2007pt}.  First, the number of internal and external lines are related via 
\be
N+2I = \sum_{i,n}i V_{(i,n)}.
\label{consedges}
\ee
Similarly, the number of internal lines is related to the number of loops in the graph via
\be
L = 1+I-\sum_{i, n}V_{(i,n)}~.
\label{defineloops}
\ee
It is convenient to use \eqref{defineloops} to eliminate $I$ from \eqref{FirstAmplitudeEstimate}, which leads to the final power-counting estimate for $\mathcal{M}^{(N)}$,
\be
{\cal M}^{(N)}(k) \sim \Lambda^{4-N}\left(\frac{ k}{\Lambda}\right)^{2L+2+\sum_{in}(n-2)V_{(i,n)}},
\label{eq:powercount}
\ee
which depends only on the number of vertices (and the number of derivatives contained therein) and the number of loops.
Using \eqref{eq:powercount}, it is easy to check what types of diagrams can renormalize coefficients in the Lagrangian, as we will see in the following sections.

The formula \eqref{eq:powercount} requires two important comments:
\begin{itemize}
\item In writing \eqref{FirstAmplitudeEstimate}, we have implicitly assumed that we are using dimensional regularization or some other mass-independent regularization scheme.  This ensures that the only scales which can emerge from loop integrals correspond to factors of external momenta.  Had we instead used a mass-dependent regularization such as Pauli--Villars or a cutoff, then factors of the regularization mass scale $\Lambda_{\rm UV}$ would also appear in \eqref{FirstAmplitudeEstimate}.  Physical results are of course regulator independent, so nothing essential is lost by using a mass-independent scheme.  Thus, unless stated otherwise, we use dimensional regularization for all calculations.

\item Logarithmic factors $\sim \log(k^{2}/\mu ^{2})$, with $\mu$ the regularization scale, are not captured by the power counting estimate. Therefore, one should think of \eqref{eq:powercount} as also potentially containing logarithmic factors when the diagrams involve loops.  Dependence on the regularization scale is only through these logarithmic factors.

\end{itemize}   

We now use this power-counting estimate \eqref{eq:powercount} to explore the behavior of various derivatively-coupled EFTs.

  \subsection{Galileons}
First, we can re-derive the non-renormalization theorem for the galileon in this language. Consider a galileon scattering process with $N$ external $\phi$ legs.   
Both the galileon operators and higher derivative terms $\sim \partial^{m}(\partial^{2}\phi)^{n}$ are included in the action.

Start by examining tree diagrams built solely from vertices drawn from the special galileon terms \eqref{GalileonTerms}.  The operator with $i$ fields has $2i-2$ derivatives, meaning that $\sum_{i,n}(n-2)V_{(i,n)}=2(N-2)$, as follows from the topological relations \eqref{consedges} and \eqref{defineloops} evaluated at $L=0$.  Therefore,
\begin{align}
\mathcal{M}^{(N)}_{\rm gal. \ tree}\sim \Lambda^{4-N}\left (\frac{k}{\Lambda}\right )^{2(N-1)}\ .\label{TreeGalileonScaling}
\end{align}
Now, we apply the estimate \eqref{eq:powercount} to all possible loop diagrams with $N$ external legs and attempt to build anything with the scaling \eqref{TreeGalileonScaling}, corresponding to a renormalization of the galileon operators.  We will find that this is not possible.
It is easy to check that loop diagrams built only from galileon vertices cannot renormalize the original operators. Using \eqref{consedges} and \eqref{defineloops} for a diagram with $L$ loops, we now have $\sum_{i,n}(n-2)V_{(i,n)}=2(N+2L-2)$, yielding the estimate
\be
\mathcal{M}^{(N)}_{\rm gal. \ loop}\sim \Lambda^{4-N}\left (\frac{k}{\Lambda}\right )^{2(N-1)+6L}\ .\label{LoopGalileonScaling}
\ee
An $L$-loop diagram thus carries $6L$ more powers of $k$ than the original tree diagrams, and therefore \textit{none} of these graphs can renormalize the original interactions.  Further, evaluating \eqref{LoopGalileonScaling} at $L=1$, it can be deduced that the higher derivative operators of the form $(\partial^{2}\phi)^{n}$ or $\partial\partial(\partial^{2}\phi)^{n}$ are not renormalized by loops of $\phi$, either. This is in agreement with explicit computations of the 1PI effective action~\cite{Luty:2003vm,Nicolis:2004qq}.

If we also use higher derivative vertices, then a vertex with $i$ external legs has at \textit{least} $2i-2$ derivatives.  This changes the relevant sum to a lower bound $\sum_{i,n}(n-2)V_{(i,n)}\ge 2(N+2L-2)$.  Such diagrams therefore have at least as many powers of $k$ as \eqref{LoopGalileonScaling}, meaning that they also will not renormalize the galileon operators.

\begin{figure}
\centering
\begin{tikzpicture}[line width=1.7 pt,baseline={([yshift=-1.2ex]current bounding box.center)},vertex/.style={anchor=base,
    circle,fill=black!25,minimum size=18pt,inner sep=2pt}]
			\draw[](135:0) -- (135:.9cm);
			\draw[](45:0) -- (45:.9cm);
			\draw[](-135:0) -- (-135:.9cm);
			\draw[](-45:0) -- (-45:.9cm);
			\node[scale=1] at (1, .05) {$\boldsymbol\sim k^{6}$};
\end{tikzpicture}					
\begin{tikzpicture}[line width=1.7 pt,baseline={([yshift=-1.2ex]current bounding box.center)},vertex/.style={anchor=base,
    circle,fill=black!25,minimum size=18pt,inner sep=2pt}]
			\draw[](135:.4) -- (135:1.2cm);
			\draw[](45:.4) -- (45:1.2cm);
			\draw[](-135:.4) -- (-135:1.2cm);
			\draw[](-45:.4) -- (-45:1.2cm);
			\node[scale=1] at (1.2, .05) {$\boldsymbol\sim k^{12}$};
			\draw[xshift=.4cm] (0,0) arc (0:360:.4);
\end{tikzpicture}				
\begin{tikzpicture}[line width=1.7 pt,baseline={([yshift=-1.2ex]current bounding box.center)},vertex/.style={anchor=base,
    circle,fill=black!25,minimum size=18pt,inner sep=2pt}]
			\draw[xshift=0cm](135:0) -- (135:.9cm);
			\draw[xshift=.8cm](45:0) -- (45:.9cm);
			\draw[xshift=0cm](-135:0) -- (-135:.9cm);
			\draw[xshift=.8cm](-45:0) -- (-45:.9cm);
			\node[scale=1] at (1.8, .05) {$\boldsymbol\sim k^{12}$};
			\draw[xshift=.8cm] (0,0) arc (0:360:.4);
\end{tikzpicture}				
\begin{tikzpicture}[line width=1.7 pt,baseline={([yshift=-1.2ex]current bounding box.center)},vertex/.style={anchor=base,
    circle,fill=black!25,minimum size=18pt,inner sep=2pt}]							
			\draw[](135:.4) -- (135:1.2cm);
			\draw[](-135:.4) -- (-135:1.2cm);
			\draw[xshift=.4cm](45:0) -- (45:.9cm);
			\draw[xshift=.4cm](-45:0) -- (-45:.9cm);
			\node[scale=1] at (1.6,.05) {$\boldsymbol\sim k^{12}$};
			\draw[xshift=.4cm] (0,0) arc (0:360:.4);
\end{tikzpicture}				
\begin{tikzpicture}[line width=1.7 pt,baseline={([yshift=-1.2ex]current bounding box.center)},vertex/.style={anchor=base,
    circle,fill=black!25,minimum size=18pt,inner sep=2pt}]
			\draw[](180:.4) -- (180:1.3cm);
			\draw[xshift=.4cm](-45:0) -- (-45:.9cm);
			\draw[xshift=.4cm](45:0) -- (45:.9cm);
			\draw[xshift=.4cm](0:0) -- (0:.9cm);
			\node[scale=1] at (2.1, .05) {$\boldsymbol\sim k^{12}$};
			\draw[xshift=.4cm] (0,0) arc (0:360:.4);
\end{tikzpicture}
\caption{\small The various $2\to 2$ diagrams built solely from galileon operators~\eqref{GalileonTerms}, up to one loop, and their scaling with external momenta. It is clear that the loop diagrams contribute at higher orders in momenta than the tree amplitude. 
}
\label{fig:2To2GalileonScatteringScaling}
\end{figure}
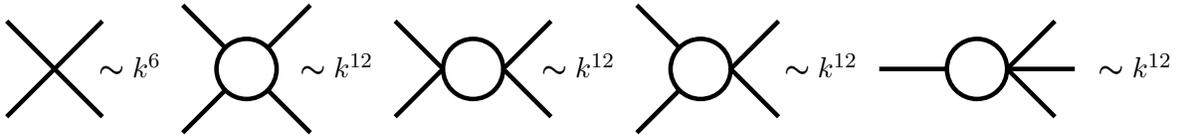

The above analysis is merely a formalization of the intuition that quickly becomes obvious when one draws diagrams.  Consider $2\to 2$ scattering to one-loop using only galileon operators, as shown in Fig. \ref{fig:2To2GalileonScatteringScaling}.  The tree diagrams built from $\mathcal{L}_{4}\sim (\partial\phi)^{2}(\partial^{2}\phi)^{2}$ or two insertions of ${\cal L}_3\sim(\partial\phi)^2\partial^2\phi$ clearly scale as $\sim k^{6}$, while the loop diagrams can be easily estimated to scale as $\sim k^{12}$.  Thus, the loops cannot renormalize the tree-level contribution. (See also \cite{Kampf:2014rka} for galileon power counting.)

Note that---in contrast to Sec.\!~\ref{Sec:NonRenormalizationTheorem}---nowhere in the preceding argument did we have to make any use of the detailed structure of the galileon interactions. Instead, the result just follows from the fact that there are certain numbers of derivatives per $\phi$ and that the galileon is massless.\footnote{If the field were massive then some of the $k$'s in \eqref{eq:powercount} could correspond to factors of mass $m$ instead of external momenta and the above analysis would not be guaranteed to work.  This is further discussed in the conclusions.} In fact, as we will see next, the galileon is not even the unique theory which has a non-renormalization theorem of this type, it is a generic property of derivatively coupled theories.

\subsection{General Relativity}
\label{sec:GRsmatrix}

As a first example, consider calculating graviton scattering diagrams in pure Einstein gravity with the cosmological constant tuned to zero, %
\begin{align}
S&=\frac{M_{\rm Pl}^{2}}{2}\int\rd^{4}x\sqrt{-g}R+\cdots\label{GRAction},
\end{align}
where $\cdots$ contains higher order operators $\sim \nabla^{m}R^{n}$. Perturbing the metric about flat space as $g_{\mu\nu}  = \eta_{\mu\nu} + {1\over M_P}h_{\mu\nu}$, the Einstein--Hilbert term is of the schematic form
\be
S \sim \int\rd^4x \sum_{n=0}^\infty \left(\frac{h}{M_{\rm Pl}}\right)^n (\partial h)^2.
\ee
Each interaction vertex now has exactly two derivatives, so that a tree diagram with $N$ external legs built from Einstein--Hilbert vertices has the scaling
\begin{align}
\mathcal{M}^{(N)}\sim M_{\rm Pl}^{4-N}\left (\frac{k}{M_{\rm Pl}}\right )^{2}.
\end{align}
 $M_{\rm Pl}$ now plays the role of $\Lambda$. In comparison, an $N$-point, $L$-loop diagram built from the Einstein--Hilbert terms scales as\footnote{Precisely the same estimate appears in DeWitt's early paper on quantum gravity~\cite{DeWitt:1967uc}.}
\begin{align}
\mathcal{M}^{(N)}\sim M_{\rm Pl}^{4-N}\left (\frac{k}{M_{\rm Pl}}\right )^{2+2L}.\label{GRPowerCounting}
\end{align}
Building loops using vertices drawn from the higher order operators contained in the $\cdots$  in \eqref{GRAction} only increases the scaling with $k$.

Therefore, we see that loops cannot correct the Einstein--Hilbert vertices: the Planck mass is not renormalized in pure flat-space GR.  Additionally, graviton loops will not cause the cosmological constant to be renormalized and so there is no cosmological constant problem in pure GR. These statements have long been  known\footnote{At one-loop, the only counterterms needed are those proportional to $R^{2}$ and to $R_{\mu\nu}^{2}$  (the other dimension-4 counterterm, proportional to $R_{\mu\nu\rho\sigma}^{2}$, is degenerate with the others by the Gauss--Bonnet theorem). This is the origin of the statement that pure GR is one-loop finite in four dimensions: the divergences all correspond to redundant operators which can be field redefined away and hence do not contribute to the S-matrix. The two-loop calculation was performed in~\cite{Goroff:1985th}, where it was found that a non-redundant counterterm $\propto R^{\mu\nu\rho\sigma}R_{\rho\sigma\kappa\lambda}R^{\kappa\lambda}{}_{\mu\nu}$ is required.  The scaling of all of these results agrees with the estimate \eqref{GRPowerCounting}.} \cite{DeWitt:1967uc,DeWitt:1985bc,'tHooft:1974bx,'tHooft:1973us}. 

\subsection{$P(X)$ Theories\label{Sec:PXTheories}} 

As the next example, consider the effective field theory of a scalar with derivative self-interactions which have at most one derivative per field. Specifically we consider Lagrangians of the form ${\cal L} = \Lambda^4P(X)$, where $X\equiv-\frac{1}{2}(\partial\phi)^{2}/\Lambda^{4}$ and $P$ is an arbitrary function. Theories of this type can be considered the leading terms in a derivative expansion of theories which possess a shift symmetry $\phi\mapsto \phi+c$. 
Consequently, they arise in myriad places, perhaps most famously as the EFT of the Nambu--Goldstone mode for a complex scalar with a symmetry breaking Mexican hat potential. $P(X)$ models have been used extensively in theoretical cosmology, both in $K$-inflation models~\cite{ArmendarizPicon:1999rj,Garriga:1999vw}, and to drive late-time acceleration in $K$-essence models~\cite{Chiba:1999ka,ArmendarizPicon:2000dh,ArmendarizPicon:2000ah}. The ghost condensate~\cite{ArkaniHamed:2003uy} is another example in this class.
In string theory, the scalar part of the action for D-branes is a special case of a $P(X)$ theory---the Dirac--Born--Infeld (DBI) model~\cite{Leigh:1989jq}. The DBI model enjoys an enhanced symmetry, $\delta\phi = x^\mu+\phi\partial^\mu\phi$, and has been used in various cosmologies~\cite{Padmanabhan:2002cp,Silverstein:2003hf,Alishahiha:2004eh,Chen:2005ad}.
Shift-symmetric models have also found application in condensed matter settings, the pure $P(X)$ theories we consider describe the effective action of superfluids~\cite{Greiter:1989qb,Son:2002zn} and suitable multi-field generalizations can describe general fluids~\cite{Dubovsky:2005xd,Dubovsky:2011sj}.

All operators in a $P(X)$ Lagrangian will have at least one derivative per $\phi$ and, after sufficient integrating by parts, can be made strictly invariant under the shift symmetry; there are no analogues of the galileon operators \eqref{GalileonTerms} for $P(X)$ theories (apart from the trivial tadpole term $\mathcal{L}\propto \phi$). 

A generic $P(X)$ action can be written as a Taylor series\footnote{Assuming that $P(X)$ is an analytic function of $X$, which will be the case if Minkowski is a sensible vacuum of the theory.  There are interesting examples where this assumption fails, e.g.,~\cite{Son:2005rv,Berezhiani:2015bqa}.}
\begin{align}
S&=\int\rd^{4}x\, \Lambda^{4}\sum_{n=1}^\infty c_{n}X^{n} +\cdots\label{GenericPofXAction},
\end{align}
where $\cdots$ contains terms with more derivatives per $\phi$.  Using \eqref{eq:powercount}, we find that a diagram with $N$ external legs built from the operators in \eqref{GenericPofXAction} scales as
\begin{align}
\mathcal{M}^{(N)}\sim \Lambda^{4-N}\left (\frac{k}{\Lambda}\right )^{N+4L}, \label{PofXCounting}
\end{align}
where we have used the fact that all the operators of interest have one derivative per field, so $\sum_{i,n}(n-2)V_{(i,n)}=\sum_{i,n}(i-2)V_{(i,n)}=N+2L-2$, after application of \eqref{consedges} and \eqref{defineloops}.

We therefore see that the contribution from loops to a given $N$-point amplitude is suppressed by positive powers of $k/\Lambda$, so the leading momentum contribution to a given amplitude is not corrected by loops. This is another example of non-renormalization: the $c_n$ coefficients in~\eqref{GenericPofXAction} cannot be changed by loops, because this would require correcting the tree level amplitudes, and loop contributions have too many powers of $k$ to do so.

Note that no part of this argument relies upon the precise form of $P(X)$, or equivalently the precise relations between the various $c_n$. We therefore see that an {\it arbitrary} function $P$ is radiatively stable in this sense. In fact, because each loop adds a factor of $k^{4}$ to the amplitude, we can actually further deduce that if we added operators of the schematic form $\mathcal{L}\sim \partial\partial X^{n}$ to the action, these would not be renormalized by loops of $\phi$, either. This is in accord with the results of~\cite{deRham:2014wfa}, who argued that an arbitrary functional form for $P(X)$ is radiatively stable---when considering self-loops and tracking only logarithmic divergences---by explicitly computing the 1-loop effective action, and that the leading corrections come with $4$ additional derivatives.

\subsection{Conformal Dilaton}
\label{sec:confgalileon}

Finally, we consider the theory of the conformal dilaton---the Goldstone of the spontaneous breaking of conformal symmetry down to Poincar\'e. In addition to the normal linear action of the Poincar\'e group, this theory is invariant under the following symmetries
\begin{align}
\delta\phi = c(1+x^{\mu}\partial_{\mu}\phi) \ ,~~~~~\delta\phi = b_{\mu}\left (2x^{\mu}+2x^{\mu}x^{\nu}\partial_{\nu}\phi-x^{2}\partial^{\mu}\phi\right )\label{ConformalGalileonSymmetry},
\end{align}
for constant $c$ and $b_\mu$, which nonlinearly realize the conformal group, SO$(4,2)$. The scalar $\phi$ also has a geometric interpretation as the small-field limit of the brane-bending mode of a Minkowski brane embedded in an Anti-de Sitter bulk~\cite{deRham:2010eu,Goon:2011qf}.\footnote{The worldvolume theory of a flat co-dimension one brane in an AdS space nonlinearly realizes the conformal group even away from the small-field limit. This nonlinear realization is actually equivalent to the parameterization we consider, with the two theories related by a complicated field redefinition~\cite{Bellucci:2002ji,Elvang:2012st,Creminelli:2013ygt}.}

This theory arises in various contexts.  It was proposed as a type of ``IR completion" of the galileon~\cite{Nicolis:2008in}, and for this reason it often goes by the name conformal galileon. It was also studied long ago by Volkov as a prototypical example of a spontaneously broken spacetime symmetry~\cite{Volkov:1973vd}. In~\cite{Sundrum:2003yt} it was argued that the conformal dilaton shares many properties with gravity (including a version of the CC problem).  It has been used to construct alternative scenarios to cosmological inflation~\cite{Rubakov:2009np,Creminelli:2010ba,Hinterbichler:2011qk,Hinterbichler:2012mv} and also plays a prominent role in the proof of the $a$-theorem in four dimensions~\cite{Komargodski:2011vj,Komargodski:2011xv}.

Despite the complicated appearance of the non-linear symmetries in \eqref{ConformalGalileonSymmetry}, it is easy to construct invariant actions for $\phi$, by simply building diffeomorphism invariant actions using the effective metric $g_{\mu\nu}=e^{2\phi}\eta_{\mu\nu}$.  The kinetic term for $\phi$ is just given by the Einstein--Hilbert term (with the ``wrong" overall sign):
\be
S_{\rm kin}=-\frac{\Lambda^{2}}{12}\int\rd^{4}x\sqrt{-g}R[g]=\int\rd^{4}x\left( -\frac{\Lambda^{2}}{2}e^{2\phi}(\partial\phi)^{2}\right) ,\label{ConformalGalileonKineticTerm}
\ee
after integrations by parts.  The free kinetic term $\sim (\partial\phi)^{2}$ is accompanied by an infinite set of specific interactions\footnote{Note that \eqref{ConformalGalileonKineticTerm} alone is a free theory in disguise, as can be seen via the field redefinition $\hat \phi=\Lambda e^{\phi}$.  When other terms are added to the action, this is no longer true, of course.\label{Foot:FreeTerm}} $\sim \phi^{n}(\partial\phi)^{2}$.  The cosmological constant term yields an exponential potential $\sqrt{-g}=e^{4\phi}$ and higher order operators are built from higher order curvature invariants made from the Riemann tensor and its covariant derivatives.

The $\mathcal{O}(R^{2})$ operators in this theory are particularly interesting and require a brief discussion. The three operators $\{R_{\mu\nu\rho\sigma}^{2},R_{\mu\nu}^{2},R^{2}\}$ are degenerate with each other and, after integrations by parts, only yield a single independent operator
\be
{\cal L}_{R^2} \propto \left[\partial^2\phi+(\partial\phi)^2\right]^2.
\label{ConformalGalileonR2Terms}
\ee
This follows the expected counting: the Gauss--Bonnet theorem removes one combination and the vanishing of the Weyl tensor removes another, resulting in the above redundancy.  

However, there also exists another four derivative operator which cannot be written in terms of four dimensional curvature invariants, is different from \eqref{ConformalGalileonR2Terms}, and is symmetric under \eqref{ConformalGalileonSymmetry}, up to a total derivative:
\be
\mathcal{L}_{\rm wz}\propto (\partial\phi)^4+2\partial^2\phi(\partial\phi)^2\label{ConformalGalileonWessZuminoTerm}.
\ee
The operator \eqref{ConformalGalileonWessZuminoTerm} is the only operator in the EFT without a four-dimensional geometric description. It has a natural interpretation as a Wess--Zumino term, hence the notation, and can be derived by coset methods applied to the breaking pattern SO$(4,2)\to$ SO$(3,1)$~\cite{Goon:2012dy}.\footnote{ A different way of deriving \eqref{ConformalGalileonWessZuminoTerm} is by constructing curvature invariants for $g_{\mu\nu}=e^{2\phi}\eta_{\mu\nu}$ in arbitrary $d$, where the Gauss--Bonnet term no longer vanishes, before taking a limit to $d\to 4$ \cite{Nicolis:2008in}.}   It is a direct analogue of the special galileon operators \eqref{GalileonTerms} and the Wess--Zumino--Witten term of the chiral Lagrangian \cite{Goon:2012dy}. Finally, the operator also appears in the flat space limit of the Wess--Zumino anomaly functional (for the $a$ anomaly of a 4D CFT)~\cite{Komargodski:2011vj,Komargodski:2011xv}.

We now apply the power counting formula \eqref{eq:powercount}. As in the General Relativity case, we will need to tune the cosmological constant term, $\sqrt{-g}=e^{4\phi}$, to zero in order to have a Poincar\'e invariant solution to expand about.  After canonically normalizing, $\phi\mapsto \phi/\Lambda$, we start by considering arbitrary diagrams built from only the kinetic term and its associated interactions \eqref{ConformalGalileonKineticTerm} (as these have the fewest derivatives).  We estimate that the $L$-loop diagram scales as 
\be
 \mathcal{M}^{(N)}_{e^{2\phi}(\partial\phi)^{2}}\sim \Lambda^{4-N}\left (\frac{k}{\Lambda}\right )^{2L+2}\label{NaiveConformalGalileonEstimate}.
\ee
From this estimate, it appears that  that one-loop diagrams constructed from \eqref{ConformalGalileonKineticTerm} alone will renormalize the 4-derivative operators \eqref{ConformalGalileonWessZuminoTerm} and \eqref{ConformalGalileonR2Terms}. However, as noted in footnote \ref{Foot:FreeTerm}, \eqref{ConformalGalileonKineticTerm} is really a free kinetic term in disguise and hence any S-matrix element constructed \textit{solely} from vertices taken from this operator will vanish, after all diagrams are summed up~\cite{Chisholm:1961tha,Kamefuchi:1961sb}, so the expression \eqref{NaiveConformalGalileonEstimate} actually vanishes.
 
 Our loop diagrams must therefore use at least one insertion of a 4-derivative \eqref{ConformalGalileonR2Terms} \eqref{ConformalGalileonWessZuminoTerm}, or higher, vertex.  Using a single four-derivative vertex and arbitrarily many vertices from the kinetic operator, we find
\be
\mathcal{M}^{(N)}\sim \Lambda^{4-N}\left (\frac{k}{\Lambda}\right )^{2L+4},
\ee
 so the first operators that can possibly be renormalized are the 6-derivative $\sim R^{3}$ and $\sim R\square R$ terms.  Replacing the 4-derivative vertex with a higher order operator or going to higher loops only increases the scaling with $k$.  Therefore, loops of $\phi$ only renormalize 6-derivative, and higher order, operators. Thus, the potential, kinetic terms and 4-derivative terms do not run in the pure conformal dilaton theory.

 \section{Coupling to Heavy Fields\label{Sec:CouplingHeavyFields}}

Given the apparent ubiquity of non-renormalization statements for derivatively coupled theories, we are led to ask: what is special about the galileon non-renormalization theorem? Specifically, the detailed argument of Sec.$\,$\ref{Sec:NonRenormalizationTheorem} would seem to be superfluous given that the same results can be derived by dimensional analysis. Does the detailed proof of non-renormalization in some way distinguish the galileon from GR, $P(X)$ and the dilaton?

We argue that indeed it does; the galileon non-renormalization theorem ensures that even coupling additional heavy fields to $\phi$ will not cause the galileon operators to be renormalized. As we will see, this is in stark contrast to the other theories we have considered, whose leading operators will generically be corrected by heavy fields.

A heuristic argument for the above claim is the following.  All of the estimates we have performed so far have assumed a mass-independent regulator.  Consider, instead, using a cutoff, Pauli--Villars or some other mass-dependent scheme.  A new scale $\Lambda_{\rm UV}$ will now arise from loop diagrams and complicate the estimates.  For instance, the 4-point loop diagram coming from two insertions of the $\sim X^{2}\sim (\partial\phi)^{4}/\Lambda^{4}$ vertex in the $P(X)$ theory will now have the schematic form
\be
\mathcal{M}^{(4)}_{X^{2},X^{2}}\sim \left (\frac{k}{\Lambda}\right)^{8}+\left (\frac{\Lambda_{\rm uv}}{\Lambda}\right )^{2}\left (\frac{k}{\Lambda}\right )^{6}+\left (\frac{\Lambda_{\rm uv}}{\Lambda}\right )^{4}\left (\frac{k}{\Lambda}\right )^{4}\label{CutoffPXEstimate}\ ,
\ee
whereas only the first term appears in dimensional regularization.
This does not change the conclusions of Sec.$\!$ \ref{Sec:PXTheories}, but just complicates the expressions.  In particular, only the first term in \eqref{CutoffPXEstimate} has a logarithmic divergence and we would again conclude that $P(X)$ loops only cause 8-derivative and higher order operators to run.  Still, we generically find power law divergences\footnote{Taking power law divergences too seriously can yield misleading conclusions, see, e.g. \cite{Burgess:1992gx}.   Logarithms provide the sharpest information, and these power law corrections should really be thought of as an estimate of the logarithmic part of the result of integrating out a heavy particle.} proportional to $k^{6}$ and $k^{4}$, corresponding to $\sim \partial\partial X^{2}$ and $\sim X^{2}$ operators which, we saw, receive no running from self-loops.

In contrast, if we used a cutoff to calculate the loops contributing to $2\to 2$ scattering in the purely galileon theory, we would find an expression of the form
\begin{align}
\mathcal{M}_{\rm gal. \ loops}^{(4)}\sim \left (\frac{k}{\Lambda}\right )^{12}+\left (\frac{\Lambda_{\rm uv}}{\Lambda}\right )^{2}\left (\frac{k}{\Lambda}\right )^{10}+\left (\frac{\Lambda_{\rm uv}}{\Lambda}\right )^{4}\left (\frac{k}{\Lambda}\right )^{8}\ .
\end{align}
There are power divergences corresponding to $\sim \partial\partial(\partial^{2}\phi)^{4}$ and $\sim (\partial^{2}\phi)^{4}$ operators (which, we found, do not run from $\phi$ loops), but no power divergence corresponding to the galileon operators \eqref{GalileonTerms}.    This is assured by the detailed non-renormalization theorem of Sec.$\!$ \ref{Sec:NonRenormalizationTheorem}: all contributions to the galileon operators, logarithmic and power law, are vanishing, regardless of the regulator.

Power laws capture, in some rough sense, the effect of coupling heavy fields to the theory.  For instance, consider pure $\lambda\phi^{4}$ theory in $d=4$,
\be
\mathcal{L}=-\frac{1}{2}(\partial\phi)^{2}-\frac{m^{2}}{2}\phi^{2}-\frac{\lambda}{4!}\phi^{4}\ .\label{MasslessPhi4}
\ee
Calculating the one-loop vacuum polarization diagram using a cutoff, one finds that there is a quadratic divergence which needs to be removed by a counterterm, $\delta m^{2}\sim \lambda \Lambda_{\rm UV}^{2}$.  Essentially, this quadratic divergence\footnote{There is also a logarithmic divergence, indicating that $m^{2}$ runs with $\beta(m^{2})\sim \lambda m^{2}$, but this is not crucial for the following discussion, so it is omitted.} indicates the generic result of coupling heavy fields to $\phi$. 

For instance, imagine that the action \eqref{MasslessPhi4} represented the leading terms in the EFT that arose from integrating out a heavy field $\Phi$ of mass $M$ which couples to $\phi$ in the following way,
\be
\mathcal{L}_{\rm UV}=-\frac{1}{2}(\partial\phi)^{2}-\frac{m^{2}}{2}\phi^{2}-\frac{1}{2}(\partial\Phi)^{2}-\frac{M^{2}}{2}\Phi^{2}-\frac{g}{2}\phi^{2}\Phi^{2}+\ldots\label{UVActionPhi4Example}\ .
\ee
The one-loop vacuum polarization diagram for $\phi$, with $\Phi$ running in the loop, now generates logarithmic running with $\beta(m^{2})\sim gM^{2}$.  This expression for the beta function is valid only for energies larger than $ \sim M$. At energies below $\Phi$'s mass, this heavy field quickly decouples and its contribution to beta functions is driven to zero; see \cite{Manohar:1996cq} for a relevant review.  So, if $m^{2}(E_{\rm UV})$ is the value of $\phi$'s mass squared parameter at a scale $E_{\rm UV}\gg M$, then its value at very low energies $E_{\rm IR}\ll M$ (denoted by $m^{2}(E_{\rm IR})$) is given by an expression of the form
\be
m^{2}(E_{\rm IR})\approx m^{2}(E_{\rm UV})+gM^{2}\log \left(\frac{E_{\rm UV}}{M}\right)\ ,\label{RelatingPhi4Masses}
\ee
up to $\mathcal{O}(1)$ factors. Above, we've ignored all other sources of running and, again, used the fact that the heavy matter decouples at energies below $M$ in order to only run $m^{2}$ between the scales $E_{\rm UV}$ and $M$ (as opposed to between $E_{\rm UV}$ and $E_{\rm IR}$, which would be incorrect). 

The above line of reasoning is an example of the sense in which power divergences demonstrate the existence of hierarchy problems.
The quadratic divergence we found in the low energy theory $\sim \lambda\Lambda_{\rm UV}^{2}$ mimics the $gM^{2}\log E_{\rm UV}/M$ correction\footnote{For $\phi$ to be active at low energies, we must have $m(E_{\rm IR})\ll M$.  This is very unnatural, as it is only true if $m^{2}(E_{\rm UV})$ lies in a very narrow window in which it can cancel off most of the $gM^{2}\log E_{\rm UV}/M$ term, otherwise $m^{2}(E_{\rm IR})\sim \mathcal{O}(M^{2})$.  A priori, there is no good reason why this should be the case, which is the usual hierarchy problem.} above with the rough correspondence $\Lambda_{\rm UV}\sim M$.  Though only logarithmic divergences are unambiguous  \cite{Burgess:1992gx}, power laws serve as acceptable proxies for how heavy physics can affect parameters in the action and the conclusions reached through either analysis are typically in agreement.

Since GR, $P(X)$ and the conformal dilaton have no non-renormalization argument which is regulator independent, we expect their amplitudes to generically have power divergences corresponding to all possible operators, hence we expect that the terms which are not renormalized by self-loops will still be sensitive to the effects of heavy fields.  We expect that the same will \textit{not} be true of galileons: coupling them to heavy fields will not renormalize the special operators \eqref{GalileonTerms}.  

The remainder of the paper is devoted to examining this claim in detail.  We couple a heavy matter field to all of the theories we discussed previously and integrate out the heavy field.  For simplicity, we couple in a non-self-interacting scalar $\Phi$ of mass $M$, being careful to use couplings which respect the symmetries of the effective theory.  We first show that the lowest-derivative galileon operators are not affected by this procedure.  We then go on to Einstein gravity, $P(X)$ theories, and the conformal dilaton in turn and demonstrate that heavy fields can affect the coefficients of all the operators of interest.  

\subsection{Integrating Out Fields via Functional Determinants\label{Sec:Method}}
In this section, we briefly introduce the formalism we employ to integrate out the heavy field, $\Phi$. For simplicity,  we only consider a non-self-interacting heavy scalar $\Phi$---coupled to the fields of interest in a manner preserving the symmetry of the EFT---so that the action is quadratic in $\Phi$. In this case, the path integral over $\Phi$ can be done exactly, and the effective action for $\phi$ is given by a functional determinant
\be
\exp iS_{\rm eff}[\phi]=\int\mathcal{D}\Phi\, \exp iS_{\phi,\Phi}= \det\left(\frac{\delta^2S_{\phi,\Phi}}{\delta\Phi\delta\Phi}\right)^{-1/2}\exp iS_{\phi,\Phi=0}\,.
\ee
Much machinery has been built to evaluate determinants of this type, for instance using heat kernels~\cite{Avramidi:2001ns,Vassilevich:2003xt}. We choose to evaluate the functional determinant perturbatively, in powers of the field $\phi$. To facilitate this, it is often best to write the determinant as a contribution to the effective action:
\be
\exp i\Delta_\Phi S_{\rm eff}=\det\left(\frac{\delta^2S_{\phi,\Phi}}{\delta\Phi\delta\Phi}\right)^{-1/2} = \exp \left(-\frac{1}{2}\Tr\log \frac{\delta^2S_{\phi,\Phi}}{\delta\Phi\delta\Phi}\right).
\ee
As a simple example, consider the two scalar action from the previous section \eqref{UVActionPhi4Example},
\begin{align}
\mathcal{L}_{\phi,\Phi}=-\frac{1}{2}(\partial\phi)^{2}-\frac{m^{2}}{2}\phi^{2}-\frac{1}{2}(\partial\Phi)^{2}-\frac{M^{2}}{2}\Phi^{2}-\frac{g}{2}\phi^{2}\Phi^{2}.
\end{align} 
The contribution to the action from integrating out $\Phi$ is then
\be
\Delta_\Phi S_{\rm eff} = \frac{i}{2}\Tr\log \frac{\delta^2S_{\phi,\Phi}}{\delta\Phi\delta\Phi} = \frac{i}{2}\Tr\log \left (\partial^{2}-M^{2}-g \phi^{2}\right ).\label{EffectiveActionExample}
\ee
Dividing through by a factor of the propagator---this shift can be absorbed into the normalization of the path integral, and corresponds to canceling the vacuum bubbles of the field $\Phi$---we can cast this as
\be
\Delta_\Phi S_{\rm eff}= \frac{i}{2}\Tr\log \left (\mathds{1}-\frac{1}{\partial^{2}-M^{2}}g \phi^{2}\right ) 
\ee

We now want to evaluate this expression perturbatively in $\phi$ by expanding the logarithm
\be
\Delta_\Phi S_{\rm eff}= -\frac{i}{2}\Tr\left(\frac{1}{\partial^{2}-M^{2}}g \phi^{2}\right)-\frac{i}{4}\Tr\left(\frac{1}{\partial^{2}-M^{2}}g \phi^{2}\frac{1}{\partial^{2}-M^{2}}g \phi^{2}\right)+\cdots.
\ee
Each term in this expression can be mapped to a particular Feynman diagram contribution.
We first evaluate the ${\cal O}(\phi^2)$ contribution to the action; in order to do this, we introduce sets of position and momentum eigenstates as outlined in Appendix~\ref{Appendix:RulesForMatrixElements} and trace over momentum eigenstates\footnote{The mapping between this situation and the notation in equation~\eqref{eq:singleinsertiontrace} is $S_{(n)}^J[ip]_J = g\phi(x)^2$, $S_{(0)}^{-1\,I}[ip]_I = -(p^2+M^2)^{-1}$.}
\begin{align}
\begin{tikzpicture}[line width=1.7 pt,baseline={([yshift=-1.2ex]current bounding box.center)},vertex/.style={anchor=base,
    circle,fill=black!25,minimum size=18pt,inner sep=2pt}]
\draw[xshift=.8cm,yshift=.42cm,dashed] (0,0) arc (0:360:.4);
\draw[style] (-.6,0) -- (1.4,0);
\end{tikzpicture}~~
&=-\frac{i}{2}\int\rd^{4}{p}\, \langle p|\frac{1}{\partial^{2}-M^{2}}g\phi^{2}|p\rangle= \frac{ig}{2}\int\rd^4 x\,\phi^2(x)\int\frac{\rd^4 p}{(2\pi)^4}\frac{1}{p^2+M^2}.
\end{align}
The (divergent) integral can then be evaluated in dimensional regularization to yield the contribution to the effective action
\begin{align}
\Delta_\Phi S_{\rm eff}&\supset -\int\rd^{4}x\, \frac{g  M^{2}}{(4\pi)^{2}}\left (\frac{1}{\epsilon}-\log M/\mu  \right )\phi(x)^{2}, 
\end{align}
where we absorbed all finite terms into the definition of $\mu$.  The $1/\epsilon$ divergence is canceled by a counterterm for the $\phi$ mass, after which we can smoothly take $\epsilon\to 0$, and the logarithm combines with the $\phi$ mass term in the full action to form
\begin{align}
S_{\rm eff}\supset \int\rd^{4}x\,-\frac{1}{2}\left (m^{2}-\frac{2g  M^{2}}{(4\pi)^2}\log M/\mu \right )\phi(x)^{2}\ .
\end{align}
Demanding that the action be independent of the arbitrary mass scale\footnote{More generally, wavefunction renormalization factors also have to be taken into account when determining the beta function, but simply demanding independence from $\mu$ will be sufficient for all the examples we consider.  The general procedure is given in, for example, \cite{GeorgiWeak}, where it is phrased in terms of the 1PI action.  The functional determinants we consider can be thought of as (part of) the one-loop contribution to the 1PI action, as calculated using the background field method \cite{Abbott:1981ke}.} $\mu$ yields the beta function $\beta(m^{2})= - \frac{2g  M^{2}}{(4\pi)^2}$, in accord with our previous expression \eqref{RelatingPhi4Masses}.  This result agrees with the standard diagrammatic analysis.

For emphasis, the beta function we just derived is only valid at energies larger than the mass of the heavy particle.  Its behavior is approximately piecewise:
\begin{align}
\beta(m^{2})\approx \begin{cases}
- \frac{2g  M^{2}}{(4\pi)^2} & E \gtrsim M\\
0 & E\lesssim M
\end{cases}\ ,
\end{align}
though the exact form of the low energy behavior is scheme dependent \cite{Manohar:1996cq}.
  The decoupling at energies below $M$ is entirely general and simply represents the fact that short distance physics has little effect on long distance physics.  In the following sections, we will derive many more beta functions and they should all be understood in the above manner, being only valid at energies larger than the heavy mass scale (which we always write as $M$).

\subsection{Galileons}

First, consider the galileon.  We couple a heavy scalar $\Phi$ to the galileon in a galileon invariant way and integrate it out.  None of the galileons are ever affected; this is for the same reason that underlies the proof of the non-renormalization theorem in Sec.$\!$ \ref{Sec:NonRenormalizationTheorem}: for the coupling between $\Phi$ and $\phi$ to be invariant, $\phi$ should only appear through $\partial^{2}\phi$ (or with more derivatives) and path integrating over $\Phi$ therefore only generates terms built strictly from $\partial^{2}\phi$.

For example, consider the case where there is a linear in $\Phi$ coupling, so that there will also be tree-level contributions to the effective action which comes from eliminating $\Phi$ via its equation of motion.  In order to see that this will not lead to galileon terms, note that there is essentially only one way to write an invariant linear coupling: $\mathcal{L}\sim f(\partial^{n}\partial^{2}\phi)\Phi$, with $f(\partial^{n}\partial^{2}\phi)$ an arbitrary scalar function of $\partial_{\mu}\partial_{\nu}\phi$ and derivatives thereof.  It can be reasoned that any other galileon invariant coupling can be put in this form after integrations by parts. Therefore, the classical EOM will take the schematic form
\begin{align}
(\partial^{2}+M^{2})\Phi=f(\partial^{n}\partial^{2}\phi)+\cdots \ ,
\end{align}
where $\cdots$ contains $\Phi$ self-interaction terms.  Then, in order to integrate out $\Phi$ at tree-level, one would merely take the original Lagrangian $\mathcal{L}(\phi,\Phi)$ and replace  $\Phi$ by
\begin{align}
\Phi\mapsto \frac{1}{M^{2}}\left (1-\frac{\partial^{2}}{M^{2}}+\ldots\right )f(\partial^{n}\partial^{2}\phi)\ .
\end{align}
This will never lead to galileons, since all $\phi$ fields have too many derivatives acting upon them.

As an example of an explicit loop computation,
take the galileon invariant coupling to $\Phi$ to be
\begin{align}
S_{\rm int}= \int\rd^4 x\left(-\frac{1}{2}(\partial\Phi)^{2}-\frac{M^{2}}{2}\Phi^{2}+ \frac{\lambda}{2\Lambda}\Phi^{2}\partial^{2}\phi \right).
\end{align}
Integrating out $\Phi$, the effective action then contains the terms
\begin{align}
\Delta_\Phi S_{\rm eff} = \frac{i}{2}\Tr\log \left(\partial^{2}-M^{2}+\frac{\lambda}{\Lambda}\partial^{2}\phi\right)\ .
\end{align}
Notice that inside the functional trace the field $\phi$ already has two derivatives acting on it.   Hence, it is already clear that no operators with fewer than two derivatives will be generated from integrating out $\Phi$.  We can check this explicitly by computing the lowest order corrections in derivatives, at order $\partial^2\phi$ we find
\be
\Delta_\Phi S_{\rm eff}\supset \int\rd^{4}x\, \frac{\lambda^{2}}{32\pi^{2}\Lambda^{2}}\left (\frac{1}{\epsilon}-\log M/\mu\right )(\partial^2\phi)^{2} + \frac{1}{192\pi^{2}M^{2}}\left(\frac{\lambda}{\Lambda}\right)^3 (\partial^2\phi)^3+\cdots,
\label{Phi2BoxpiResult}
\ee
where $\cdots$ indicates terms which are both higher order in fields with 2 derivatives per field and terms with more derivatives per field. We see from this cubic action that in particular the cubic galileon is not generated. This continues to be true at higher order---only terms with at least two derivates per field are generated.

Again, these results can be seen to follow from the fact that the interaction always involves a power of $\partial^{2}\phi$ which always results in effective actions built from $\partial^{2}\phi$.  A more interesting case would be if there were a Wess--Zumino like coupling between $\Phi$ and $\phi$ which was invariant under \eqref{GalileonSymmetryIntro} up to a total derivative but could not be integrated by parts into a form where there are at least two derivatives per $\phi$, however we are not aware of any such couplings. Note that if the coupling is not invariant, for example the standard $\sim \phi T$ matter coupling often considered in studies of the Vainshtein mechanism, then these non-renormalization statements do not strictly hold \cite{Heisenberg:2014raa}.

We've been somewhat agnostic about the role of the heavy $\Phi$ field, but one might have hoped that it could play some significant part in the UV completion of the non-renormalizable galileon theory.  The above calculations then provide explicit evidence against the possibility of UV completing galileons in a standard, local manner, in accordance with the general arguments of \cite{Adams:2006sv} against any such completion.

\subsection{General Relativity\label{GRsectionf}}

Now we will see that the non-renormalization of the Planck mass and cosmological constant of Sec.\!~\ref{sec:GRsmatrix} does not hold upon integrating out a heavy field.  Consider coupling minimally a heavy scalar, $\Phi$, to GR,
 \begin{align}
 S&=\int\rd^{4}x\sqrt{-g}\, \left (\frac{M_{\rm Pl}^{2}R}{2}-\Lambda+\frac{1}{2}\Phi(\square- M^{2})\Phi\right ).
 \end{align}
 The result of integrating out $\Phi$ is well known (see e.g. \cite{Birrell:1982ix}, or \cite{Shapiro:1999zt,Martin:2012bt}).  There is a contribution to the cosmological constant which can be computed from the zero momentum part of the diagram
\be \begin{tikzpicture}[line width=1.7 pt,baseline={([yshift=-.5ex]current bounding box.center)},vertex/.style={anchor=base,
    circle,fill=black!25,minimum size=18pt,inner sep=2pt}]
\draw[xshift=.8cm] (0,0) arc (0:360:.4);
\draw[style={decorate,decoration=complete sines},line width=1] (0,-.05) -- (-1.5,-.05);
\draw[style={decorate,decoration=complete sines},line width=1] (0,.05) -- (-1.5,.05);
\end{tikzpicture}~.
\ee
The contribution to the effective action is
  \begin{align}
S_{\rm eff}&\supset \int\rd^{4}x\, \sqrt{-g}\left [-\Lambda+\frac{M^{4}}{32\pi^{2}}\left (\frac{1}{\epsilon}-\log M/\mu \right )\right ].
\end{align} We introduce the renormalized CC as $\Lambda=\Lambda_{R}+\delta\Lambda$ with $\delta \Lambda=\frac{1}{\epsilon}\frac{M^{4}}{32\pi^{2}}$, take $\epsilon\to 0$ safely and then demand that $S_{\rm eff}$ be independent of $\mu $ to yield the beta function for $\Lambda$,
\begin{align}
\beta(\Lambda)&=\frac{M^{4}}{32\pi^{2}}\ .\label{CCbeta}
\end{align}
 
 There is a contribution to the Planck mass which can be computed from the order $\partial^2$ part of the diagram
 \be
 \begin{tikzpicture}[line width=1.7 pt,baseline={([yshift=-.5ex]current bounding box.center)},vertex/.style={anchor=base,
    circle,fill=black!25,minimum size=18pt,inner sep=2pt}]
\draw[xshift=.8cm] (0,0) arc (0:360:.4);
\draw[style={decorate,decoration=complete sines},line width=1] (0,-.05) -- (-1.3,-.05);
\draw[style={decorate,decoration=complete sines},line width=1] (0,.05) -- (-1.3,.05);
\draw[xshift=.8cm,style={decorate,decoration=complete sines},line width=1] (0,-.05) -- (1.3,-.05);
\draw[xshift=.8cm,style={decorate,decoration=complete sines},line width=1] (0,.05) -- (1.3,.05);
\end{tikzpicture}
\ee
 which leads to a contribution to the effective action of the form
 \be
S_{\rm eff} \supset \int\rd^{4}x \sqrt{-g}\left[M_{\rm Pl}^2-\frac{M^2}{48\pi^2}\left(\frac{1}{\epsilon}-\log M/\mu\right)\right]\frac{R}{2}+\cdots.
\ee

In order to cancel off the pole, we introduce a renormalized Planck mass and associated counterterm. Then, demanding that the answer is independent of the renormalization scale, $\mu$, yields the beta function for the Planck mass
\be
  \beta(M_{\rm Pl}^{2})=\frac{M^{2}}{48\pi^{2}}\ .\label{BetaFunctionMpScalar}
\ee

  \subsection{$P(X)$ Theories}
  
  In this section we couple $\Phi$ to $P(X)$ theories, and integrate it out to see to the renormalization of $P(X)$.  We do this in several different ways.

\subsubsection{$\partial_{\mu}\phi\partial_{\nu}\phi T^{\mu\nu}$ Coupling \label{sec:TmunuSection} }

 The couplings we consider must respect the $\phi\mapsto \phi+c$ symmetry of $P(X)$ theories. One coupling which satisfies this criterion is coupling the stress tensor of the $\Phi$ field to derivatives of $\phi$, as might arise from certain brane-world constructions, 
\begin{align}
\mathcal{L}&= \Lambda^4 P(X)-\frac{1}{2}(\partial\Phi)^{2}-\frac{M^{2}}{2}\Phi^{2}+ \frac{\lambda}{\Lambda^{4}}\partial_{\mu}\phi\partial_{\nu}\phi T^{\mu\nu}(\Phi).\label{TmunuCoupling}
\end{align}
Here the stress tensor for $\Phi$ is
\be
T_{\mu\nu}(\Phi) = \partial_{\mu}\Phi\partial_{\nu}\Phi-\eta_{\mu\nu}\left (\frac{1}{2}(\partial\Phi)^{2}+\frac{M^{2}}{2}\Phi^{2}\right). 
\ee
Integrating out $\Phi$, the effective action then contains the generated terms
\be
\Delta_\Phi S_{\rm eff} = \frac{i}{2}\Tr\log \left (\square-M^{2} -\frac{\lambda M^{2}}{\Lambda^{4}}(\partial\phi)^2-\frac{2\lambda}{\Lambda^4}\square\phi\partial^\mu\phi\partial_\mu+\frac{\lambda}{\Lambda^4}(\partial\phi)^2\square - \frac{2\lambda}{\Lambda^4}\partial^\mu\phi\partial^\nu\phi\partial_\mu\partial_\nu\right).
\label{TheFullTrace}
\ee
The simplest possible computation we can do is to check if this coupling to the heavy field $\Phi$ induces a wavefunction renormalization of the kinetic term $\sim-\frac{1}{2}(\partial\phi)^2$. The only operator that can contribute to this is the $-\frac{\lambda M^{2}}{\Lambda^{4}}(\partial\phi)^2$ term in~\eqref{TheFullTrace}, so we just have to compute the single insertion trace involving this operator:
\begin{align}
\begin{tikzpicture}[line width=1.7 pt,baseline={([yshift=-1.2ex]current bounding box.center)},vertex/.style={anchor=base,
    circle,fill=black!25,minimum size=18pt,inner sep=2pt}]
\draw[xshift=.8cm,yshift=.42cm,dashed] (0,0) arc (0:360:.4);
\draw[style] (-.6,0) -- (1.4,0);
\end{tikzpicture}~~= -\frac{i}{2}\Tr\left( \frac{1}{\partial^2 - M^2} \frac{\lambda M^2}{\Lambda^4}(\partial\phi)^2\right) = \int\rd^4 x-\frac{1}{2}(\partial\phi)^2 \frac{\lambda M^{4}}{(4\pi)^{2}\Lambda^4}\left (\frac{1}{\epsilon}-\log M/\mu  \right ).
\end{align}
We see that the $P(X)$ kinetic term \textit{does} get renormalized by the heavy field, i.e., an anomalous dimension\footnote{The anomalous dimension is $\gamma_{\phi}=\frac{\lambda M^{2}}{2(4\pi)^{2}\Lambda^{2}}$, as can be derived using eq. (1a.1.39) of \cite{GeorgiWeak}.} is acquired, when we couple $\phi$ to $\Phi$ as in \eqref{TmunuCoupling}.  In comparison, loops of $\phi$ in a pure $P(X)$ theory do not induce such an anomalous dimension.  

\subsubsection{DBI \label{Sec:DBI}}

Next, we consider a special case of $P(X)$ theories: the DBI Lagrangian. This theory describes the dynamics of a Minkowskian 3-brane embedded in a five dimensional spacetime.  The action is built through the induced metric on the brane,
\be
\bar{g}_{\mu\nu}=\eta_{\mu\nu}+\frac{1}{\Lambda^4}\partial_{\mu}\phi\partial_{\nu}\phi\  \ , \label{induceddbimetrice}
\ee
 and its associated curvature invariants, including the extrinsic curvature tensor $K_{\mu\nu}=-\gamma \partial_{\mu}\partial_{\nu}\phi$.  In particular, the DBI kinetic term comes from the volume element:
\be
S_{\rm DBI}=-\Lambda^{4}\int\rd^{4}x\sqrt{-\bar{g}}=-\Lambda^{4}\int\rd^{4}x\sqrt{1+\frac{(\partial\phi)^{2}}{\Lambda^4}}\ . \label{DBIAction}
\ee

In addition to the $\phi\mapsto\phi+c$ symmetry of all $P(X)$ theories, DBI is further symmetric under $\phi \mapsto \phi+b_{\mu}(x^{\mu }+\phi\partial^{\mu}\phi)$. In this context, the $P(X)$ symmetry is the worldvolume consequence of higher-dimensional translation invariance of the brane along the transverse direction, while the second symmetry is a consequence of higher-dimensional boosts mixing brane directions with the transverse direction.

Couplings to a heavy scalar which are DBI-invariant are easy to engineer by utilizing the induced metric $\bar g_{\mu\nu}$ in \eqref{induceddbimetrice}. The simplest such coupling takes the form:
\be
S_{\rm coupling}=\int\rd^{4}x\sqrt{-\bar{g}}\left (-\frac{1}{2}\bar{g}^{\mu\nu}\partial_{\mu}\Phi\partial_{\nu}\Phi-\frac{M^{2}}{2}\Phi^{2}\right )\ .\label{DBIMatterCoupling}
\ee
We would like to understand how the presence of this heavy scalar renormalizes the tension $\Lambda$. To 
calculate this, we will determine the contributions to the $\sim(\partial\phi)^{2}$ and $\sim(\partial\phi)^{4}$ terms 
in the effective action, and verify that they match the expansion in \eqref{DBIAction}.  

The functional determinant we want to compute is (again, we drop the contribution from $\sqrt{-\bar g}$)
\be
\Delta_\Phi S_{\rm eff} = \frac{i}{2}\log\det\left( \bar{\square}-M^{2}\right) = \frac{i}{2}\Tr\log\left( \bar{\square}-M^{2}\right)
\ee
with $\bar{\square} = \bar g^{\mu\nu}\bar\nabla_\mu\bar\nabla_\nu$ built from covariant derivatives with respect to $\bar{g}_{\mu\nu}$.  
The easiest way of computing $\bar{\square}$ is to use
\begin{align}
\bar{\square}=&\,\frac{1}{\sqrt{-\bar{g}}}\partial_{\mu}\left [\bar{g}^{\mu\nu}\sqrt{-\bar g}\partial_{\nu}\right ]\nn
=&\,\partial^2-\frac{1}{\Lambda^4}\partial^2\phi\partial_{\mu}\phi\partial^{\mu}- \frac{1}{\Lambda^4}\partial^{\mu}\phi\partial^{\nu}\phi\partial_{\mu}\partial_{\nu}\nn
&+\frac{1}{\Lambda^8}\partial^2\phi(\partial\phi)^2\partial^\mu\phi\partial_\mu+\frac{1}{2\Lambda^8}\partial^\mu(\partial\phi)^2\partial_\mu\phi\partial^\nu\phi\partial_\nu+\frac{1}{\Lambda^8}(\partial\phi)^2\partial^\mu\phi\partial^\nu\phi\partial_\mu\partial_\nu\cdots
\label{DBIBoxOperatorExpansion} ,
\end{align}
where we have used $\bar{g}^{\mu\nu}=\eta^{\mu\nu}-\frac{\gamma^{2}}{\Lambda^4}\partial^{\mu}\phi\partial^{\nu}\phi \ , \gamma\equiv 1/\sqrt{1+\frac{(\partial\phi)^{2}}{\Lambda^4}}$,
and indices are raised and lowered with $\eta_{\mu\nu}$.  We are only looking for terms which can generate terms $\sim(\partial\phi)^{2}$ or $\sim(\partial\phi)^{4}$, and the pieces in \eqref{DBIBoxOperatorExpansion} which have two derivatives acting on a $\phi$ cannot give rise to these. The relevant trace is then reduced to
 \begin{align}
\Delta_\Phi S_{\rm eff}&=\frac{i}{2}\Tr\log \left(\partial^{2}-M^{2}- \frac{1}{\Lambda^4}\partial^{\mu}\phi\partial^{\nu}\phi\partial_{\mu}\partial_{\nu}+ \frac{1}{\Lambda^8}(\partial\phi)^{2}\partial^{\mu}\phi\partial^{\nu}\phi\partial_{\mu}\partial_{\nu}\right).
 \end{align}

The contribution to the $\sim(\partial\phi)^{2}$ term is given by a single insertion trace over the  $- \partial^{\mu}\phi\partial^{\nu}\phi\partial_{\mu}\partial_{\nu}$ operator:
 \be
 \begin{tikzpicture}[line width=1.7 pt,baseline={([yshift=-1.2ex]current bounding box.center)},vertex/.style={anchor=base,
    circle,fill=black!25,minimum size=18pt,inner sep=2pt}]
\draw[xshift=.8cm,yshift=.42cm,dashed] (0,0) arc (0:360:.4);
\draw[style] (-.6,0) -- (1.4,0);
\end{tikzpicture}~~= -\frac{i}{2}\Tr\left( \frac{1}{\partial^2 - M^2} \frac{1}{\Lambda^4}\partial^{\mu}\phi\partial^{\nu}\phi\partial_{\mu}\partial_{\nu}\right)= \int\rd^{4}x\, \frac{1}{2}(\partial\phi)^{2} \frac{M^{4}}{32\pi^{2}\Lambda^4}\left (\frac{1}{\epsilon}-\log M/\mu \right ). \label{LinearDBITrace}
 \ee
 and, summing up both a single and double insertion trace, the $\sim(\partial\phi)^{4}$ terms are given by
\begin{align}
 \begin{tikzpicture}[line width=1.7 pt,baseline={([yshift=-.6ex]current bounding box.center)},vertex/.style={anchor=base,
    circle,fill=black!25,minimum size=18pt,inner sep=2pt}]
\draw[](120:0) -- (120:.7cm);
\draw[](160:0) -- (160:.7cm);
\draw[](200:0) -- (200:.7cm);
\draw[](240:0) -- (240:.7cm);
\draw[xshift=.8cm,dashed] (0,0) arc (0:360:.4);
\end{tikzpicture}~+
 \begin{tikzpicture}[line width=1.7 pt,baseline={([yshift=-.6ex]current bounding box.center)},vertex/.style={anchor=base,
    circle,fill=black!25,minimum size=18pt,inner sep=2pt}]
\draw[](135:0) -- (135:.7cm);
\draw[xshift=.8cm](45:0) -- (45:.7cm);
\draw[](-135:0) -- (-135:.7cm);
\draw[xshift=.8cm](-45:0) -- (-45:.7cm);
\draw[xshift=.8cm,dashed] (0,0) arc (0:360:.4);
\end{tikzpicture}=
\int\rd^{4}x-\frac{1}{8\Lambda^4}(\partial\phi)^{4} \frac{M^{4}}{32\pi^{2}\Lambda^4}\left (\frac{1}{\epsilon}-\log M/\mu \right )\ .
\end{align}
The  relative coefficient matches precisely what we obtain from the expansion of the volume element:
\begin{align}
\Lambda^4\sqrt{-\bar{g}}\approx 1+\frac{1}{2}(\partial\phi)^{2}-\frac{1}{8\Lambda^4}(\partial\phi)^{4}+\ldots
\end{align}
Comparing to \eqref{DBIAction} the renormalized tension $\Lambda$ is seen to run as
\begin{align}
\beta(\Lambda^{4})&=\frac{M^{4}}{32\pi^{2}}\ .\label{DBITensionRenormalization}
\end{align}
Therefore, we explicitly see that a heavy scalar renormalizes the infinite tower of $\sim X^{n}$ operators which appear in the DBI action.

Note that we have essentially just repeated the calculation of the CC running of section \ref{GRsectionf}: $\Phi$ couples to $\phi$ via minimal coupling to $\bar{g}_{\mu\nu}$,  which is just a specific choice of metric.  It is therefore not surprising that the two beta functions~\eqref{DBITensionRenormalization} and~\eqref{CCbeta} agree.

\subsection{Conformal Dilaton}

Finally, we couple $\Phi$ to the conformal dilaton field, denoted by $\phi$. 
We take $\Phi$ to transform as a primary field of weight $\Delta$ so that under the conformal symmetries $\Phi$ transforms as
 \begin{align}
\delta\Phi = c\left(\Delta+x^\mu\partial_\mu\right)\Phi\,,~~~~~~~
\delta\Phi = b_{\mu}\left (2\Delta x^\mu+2x^{\mu}x^{\nu}\partial_{\nu}-{x^{2}}{}\partial^{\mu}\right )\Phi \ ,\label{MatterConfGalSymmetries}
\end{align}
with constant $c,b_\mu$, while the dilaton transforms non-linearly as in equation~\eqref{ConformalGalileonSymmetry}.

After fixing the mass and scaling dimension of $\Phi$ and canonically normalizing, there exists a one-parameter class of two derivative interactions which are quadratic in $\Phi$ and symmetric under \eqref{MatterConfGalSymmetries} and \eqref{ConformalGalileonSymmetry}:
\begin{align}
S_{\rm int}&=\int\rd^{4}x\, -\frac{e^{2\phi(1-\Delta)}}{2}(\partial\Phi)^{2}-\frac{M^{2}e^{2\phi(2-\Delta)}}{2}\Phi^{2}-\left (\lambda+\Delta-2\lambda\Delta-\Delta^{2}\right )\frac{e^{2\phi(1-\Delta)}}{2}\Phi^{2}(\partial\phi)^{2}\nn
&\quad -\lambda e^{2\phi(1-\Delta)}\Phi\partial^{\mu}\Phi\partial_{\mu}\phi\ .
\end{align}
The $\Delta=1$, $\lambda=0$ case was considered in~\cite{Komargodski:2011vj}, as a check of their arguments in the proof of the $a$-theorem.

Working to fourth order in derivatives, we integrate out $\Phi$ and find, after a lengthy calculation:
\begin{align}
\Delta_\Phi S_{\rm eff} &=\int\rd^{4}x\, \left [d_{V}e^{4\phi}-d_{R}e^{2\phi}(\partial\phi)^{2}+d_{R^{2}}\left (\square\phi+(\partial\phi)^{2}\right )^{2}\right ]\left (\frac{1}{\epsilon}-\log M/\mu \right)\nn
&\quad+\int\rd^{4}x\, \left [f_{V}e^{4\phi}-f_{R}e^{2\phi}(\partial\phi)^{2}+f_{R^{2}}(\square\phi+(\partial\phi)^{2})^{2}+f_{\rm WZ}\left (\square\phi+2\square\phi(\partial\phi)^{2}\right )\right ]\ ,
\end{align}
where the coefficients of the divergent and finite terms are:
\begin{align}
\begin{pmatrix}
d_{V}\\
d_{R}\\
d_{R^{2}}\\
f_{V}\\
f_{R}\\
f_{R^{2}}\\
f_{\rm WZ}
\end{pmatrix}&= \frac{1}{(4\pi)^{2}}\begin{pmatrix}
\frac{M^{4}}{2}\\
-M^2 (\Delta +\lambda -1),\\ 
\frac{1}{2} (\Delta
   +\lambda -1)^2\\
\frac{3M^{4}}{8}\\
-\frac{M^{2}}{3}\\
\frac{1}{60}+ \frac{(\Delta+\lambda) (\Delta+\lambda-1)}{3}\\
   -\frac{1}{180}
\end{pmatrix}\label{MostGeneralConformalCoefficients}\ .
\end{align}
In order to consistently calculate \eqref{MostGeneralConformalCoefficients} using dimensional regularization, one must also work with the invariant operators $\sqrt{-g}$ , $\sqrt{-g}R$, etc. (where, again, $g_{\mu\nu}=e^{2\phi}\eta_{\mu\nu}$) as constructed in arbitrary dimensions.  This induces additional factors of $\epsilon$'s, as in $\sqrt{-g}=e^{d\phi}=e^{(4-\epsilon)\phi}$, which contribute non-trivially to the finite parts of the functional determinant. Such terms are necessary in obtaining a conformally invariant answer. 

We see that the Wess-Zumino term receives a finite renormalization.  The coefficient of the Wess--Zumino term, $f_{\rm WZ}$, is notably independent of both $\lambda$ and $\Delta$, as it should be, since $f_{\rm WZ}$ is directly related to the $a$-anomaly which characterizes a fundamental property of the free scalar field which shouldn't depend on how one couples the dilaton to $\Phi$.  Its numerical value is in full agreement with the result in~\cite{Komargodski:2011vj}.

\section{Conclusions\label{Sec:Conclusion}}

Many massless, derivatively coupled effective theories have non-renormalization theorems that follow simply from dimensional analysis, without reference to the detailed structure of the interactions.  Essentially, their interactions have so many derivatives that the self-loops can only affect operators of a certain minimum dimension.  Among the theories in this class are General Relativity, $P(X)$ theories, the conformal dilaton (sometimes called the conformal compensator, or conformal galileon), and galileons. 

Galileons, however, possess a stronger, diagrammatic non-renormalization theorem that depends on the detailed structure of the galileon interaction terms \cite{Hinterbichler:2010xn}.  We have interpreted the extra strength of the galileon non-renormalization theorem as the statement that the galileon operators are not renormalized even by loops of {other} heavy fields, as long as they are coupled in a galileon invariant way.  In the other theories like GR or $P(X)$, the leading operators {are} renormalized by these heavy loops. 

 We have tested this interpretation by coupling a heavy scalar field to each of these theories in ways that respect the symmetries of each theory.  None of the five special galileon operators are affected by heavy fields. In comparison, the operators in the GR, $P(X)$ and conformal dilaton theories which were not affected by self-loops \textit{are} renormalized by loops of the heavy field.

Finally, we note that even if the galileon symmetry is not an exact symmetry but is weakly broken, the non-renormalization theorem still controls corrections to the galileon terms, rendering them proportional to the small breaking, a fact which can be useful, for example in constructing technically natural cosmological models \cite{Pirtskhalava:2015nla,Gratia:2016tgq}.

As an example of this, we note another difference in the non-renormalization theorem of galileons compared with other theories that becomes apparent when we consider deforming the theory with a mass.  In the power counting formula \eqref{eq:powercount}, we were able to find the scaling of arbitrary diagrams with powers of the scale $\Lambda$ and the external momenta of the diagram $k$.  When the theory is massless, $k$ truly refers to an external momenta, but if there are massive particles running in the amputated diagram, then factors of $k$ can \textit{also} represent mass scales.  For instance, consider adding a mass term to the leading operators in a $P(X)$ theory:
\be
\mathcal{L}=-\frac{1}{2}(\partial\phi)^{2}-\frac{m^{2}}{2}\phi^{2}-\frac{\lambda}{\Lambda^{4}}(\partial\phi)^{4}+\ldots\ .
\ee
The power counting formula \eqref{eq:powercount} still tells us that the tree and one-loop $2\to 2$ diagrams scale as
\be
\mathcal{M}^{(4)}_{(\partial\phi)^{4}}\sim \left (\frac{k}{\Lambda}\right )^{4} \ , \quad \mathcal{M}^{(4)}_{(\partial\phi)^{4},(\partial\phi)^{4}}\sim \left (\frac{k}{\Lambda}\right )^{8} \ , 
\ee
respectively.  However, the meaning is different now, since some of the $k$'s can correspond to mass scales and so in addition to terms like $\left (\frac{k}{\Lambda}\right )^{8}$ we can also have terms like $\sim \left (\frac{m}{\Lambda}\right )^{4}\left (\frac{k}{\Lambda}\right )^{4}$.  This would correspond to a running of $\lambda$ with $\beta(\lambda)\sim \lambda^{2}\left (\frac{m}{\Lambda}\right )^{4}$. 

Therefore, we see that generically adding mass scales ruins the analyses in section \ref{Sec:DimensionalArguments}. However, there is an exception with the galileons. Though it would seem that powers of $k^{2}$ could turn into factors of $m^{2}$ in the power counting formulae and result in running galileon couplings, the detailed version of the non-renormalization theorem ensures that this does not occur.  This can be seen from the path integral proof in Sec.$\!$ \ref{Sec:NonRenormalizationTheorem}; adding a mass term changes nothing about the argument.  This is yet another way that the galileons are different from GR, $P(X)$ and the conformal dilaton: the non-renormalization theorem survives when the theory is deformed by a mass \cite{Burrage:2010cu}.

\paragraph{Acknowledgments:}
We thank Brando Bellazzini for helpful correspondence.
Research at Perimeter Institute is supported by the Government of Canada through Industry Canada and by the Province of Ontario through the Ministry of Economic Development and Innovation.  This work was made possible in part through the support of a grant from the John Templeton Foundation. The opinions expressed in this publication are those of the author and do not necessarily reflect the views of the John Templeton Foundation (KH). This work was supported in part by the Kavli Institute for Cosmological Physics at the University of Chicago through grant NSF PHY-1125897, an endowment from the Kavli Foundation and its founder Fred Kavli, and by the Robert R. McCormick Postdoctoral Fellowship (AJ). The work of M.T. was supported in part by US Department of Energy (HEP) Award DE-SC0013528. GG gratefully acknowledges support from a Starting Grant of the European Research Council (ERC StG grant 279617).

\appendix

\section{Evaluating Functional Determinants\label{Appendix:RulesForMatrixElements}}

In this Appendix, we review the procedure for evaluating the traces in the functional determinants. For another recent application of this general technique, see~\cite{Henning:2016lyp}. The general object we are interested in computing is the 1-loop correction to the action
\be
\Delta\Gamma_{(1)} = \frac{i}{2}\log \det\left(\frac{\delta^2 S}{\delta\phi^i\delta\phi^j}\right)= \frac{i}{2}\Tr\log \left(\frac{\delta^2 S}{\delta\phi^i\delta\phi^j}\right),
\ee
where the $\phi^i$ are the fields in the theory. Defining $\frac{\delta^2 S}{\delta\phi^i\delta\phi^j} \equiv S_{ij}$ we can expand in powers of $\phi^i$ as
\be
S_{ij} = S_{(0)ij}+S_{(1)ij}+S_{(2)ij}\cdots,
\ee
after factoring out  $S_{0ij}$ (which is field independent), we can expand the logarithm %
to obtain
\begin{align}
\label{eq:Straces}
\Delta\Gamma_{(1)}  &= \frac{i}{2}\Tr\log {\bf S}_{(0)}\left({\mathds 1}+{\bf S}_{(0)}^{-1}{\bf S}_{(1)}+{\bf S}_{(0)}^{-1}{\bf S}_{(2)}+\cdots\right)\\\nonumber
&= \frac{i}{2}\Tr\log {\bf S}_{(0)}+\frac{i}{2}\Tr\left({\bf S}_{(0)}^{-1}{\bf S}_{(1)}\right)+\frac{i}{2}\Tr\left({\bf S}_{(0)}^{-1}{\bf S}_{(2)}\right)+\frac{i}{2}\Tr\left({\bf S}_{(0)}^{-1}{\bf S}_{(1)}{\bf S}_{(0)}^{-1}{\bf S}_{(1)}\right)+\cdots,
\end{align}
where we have employed matrix notation ${\bf S}_{(n)} \leftrightarrow S_{(n)ij}$. The piece $\Tr\log {\bf S}_{(0)}$ is independent of the fields in the action, and just represents a constant shift of the vacuum energy of the theory, so we will often discard it.

In order to evaluate the traces in~\eqref{eq:Straces}, it is useful to make a quantum-mechanical analogy, the operators whose trace we want to evaluate are local functions of coordinates and derivatives: $S_{(n)ij}(x, \partial)$. We therefore promote the coordinates and derivatives to operators acting on a Hilbert space as
\be
\label{eq:xandDtooperators}
x^\mu \mapsto \hat x^\mu\,,~~~~~~~~~\partial_\mu\mapsto i\hat p_\mu.
\ee
We then introduce sets of position eigenstates, $\lvert x\rangle$ and momentum eigenstates $\lvert p\rangle$, which satisfy the orthogonality and completeness relations
\begin{align}
&\langle x\rvert y\rangle = \delta(x-y)~~~~~~~~~~~~~~~~\langle p\rvert k\rangle = \delta(p-k)\\
&\int\rd^dx\,\lvert x\rangle\langle x\rvert = \mathds{1}~~~~~~~~~~~~~~\int\rd^dp\,\lvert p\rangle\langle p\rvert = \mathds{1}.
\end{align}
The inner product between these two bases is given by
\be
\langle x\lvert p\rangle = \frac{1}{(2\pi)^{d/2}}e^{i p\cdot x}.
\ee
The action of the $\hat x^\mu$ and $\hat p_\mu$ operators on these eigenstates is the obvious one
\begin{align}
\hat x^\mu\lvert x\rangle &= x^\mu\lvert x\rangle\\
\hat p_\mu\lvert p\rangle &= p_\mu\lvert p\rangle.
\end{align}
Any local operator ${\cal O}(x, \partial)$ is of the form
\be
{\cal O}(x, \partial) = {\cal O}(x)+{\cal O}^{\mu_1}(x)\partial_{\mu_1}+{\cal O}^{\mu_1\mu_2}(x)\partial_{\mu_1}\partial_{\mu_2}+\cdots \equiv {\cal O}^I(x)\partial_I,
\ee
where $I$ is a multi-index. The coefficients ${\cal O}^{\mu_1\mu_2\cdots}(x)$ in this expression are built out of the fields $\phi^i$ and their derivatives. To evaluate the traces, we make the replacement~\eqref{eq:xandDtooperators}
\be
{\cal O}(x, \partial) \mapsto  {\cal O}^I(\hat x)[i\hat p]_I.
\ee
We will evaluate the traces in momentum space, so we want to evaluate the matrix element of this operator between momentum eigenstates
\begin{align}
\nonumber
\langle k\rvert {\cal O}^I(\hat x)[i\hat p]_I\rvert p\rangle &= \int\rd^dx\,\langle k\rvert {\cal O}^I(\hat x)\lvert x\rangle \langle x\rvert[i\hat p]_I\rvert p\rangle = \int\rd^dx\,{\cal O}^I( x)[ip]_I\langle k\rvert x\rangle \langle x\rvert p\rangle\\
&= \int\rd^dx\, e^{-i(k-p)\cdot x}(2\pi)^{-d}{\cal O}^I( x)[ip]_I = (2\pi)^{-d}\tilde{\cal O}^I(k-p)[ip]_I,
\label{eq:Omomentummatrix}
\end{align}
where tilde refers to the Fourier transform\footnote{We employ the following convention for Fourier transformation
\be
f(x) = \int\frac{\rd^dk}{(2\pi)^d}\,e^{ik\cdot x}\tilde f(k)~~~~~~~~\tilde f(k) = \int \rd^dx\,e^{-ik\cdot x}f(x).
\ee}.
with this rule we can evaluate any trace that we encounter.

\paragraph{Propagator:} The matrix element of the operator ${\bf S}_{(0)}(\partial)=S_{(0)ij}^I[i\hat p]_I$, with all the $S_{(0)ij}^I = {\rm const.}$ can be evaluated as
\be
\langle k\rvert S^I_{(0)ij}[i\hat p]_I\rvert p\rangle =S^I_{(0)ij}[i p]_I\delta(p-k).
\ee
The matrix elements for the propagator ${\bf S}_{(0)}^{-1}$ can be evaluated similarly:
\be
\langle k\rvert S^{-1\,I}_{(0)ij}[i\hat p]_I\rvert p\rangle =S^{-1\,I}_{(0)ij}[i p]_I\delta(p-k).
\label{eq:invpropmatrix}
\ee

\paragraph{Single insertion trace:} Frequently, we will want to compute the trace involving an insertion of a single operator of order $n$ in the background fields. This takes the form
\begin{align}
\frac{i}{2}\Tr\left({\bf S}_{(0)}^{-1}{\bf S}_{(n)}\right) &= \frac{i}{2}\int\rd^dp\,\langle p\rvert S_{(0)}^{-1\,I\,ij}[i\hat p]_I\,S_{(n)ij}^{J}(\hat x)[i\hat p]_J\lvert p\rangle\\
&= \frac{i}{2} \tilde S^J_{(n)ij}(0)\int\frac{\rd^dp}{(2\pi)^d} [ip]_J S_{(0)}^{-1\,I\,ij}[ ip]_I,
\end{align}
where to get to the second line we have inserted a complete set of momentum eigenstates and used the formulae~\eqref{eq:Omomentummatrix} and \eqref{eq:invpropmatrix}. The factor $\tilde S^J_{(n)jk}(0)$ seems at first sight somewhat strange, but we can rewrite the Fourier transform at zero momentum as an integral over all of position space to obtain the form
\be
\label{eq:singleinsertiontrace}
\frac{i}{2}\Tr\left({\bf S}_{(0)}^{-1}{\bf S}_{(n)}\right)  = \frac{i}{2}\int\rd^d x\,S^J_{(n)ij}(x) \int\frac{\rd^dp}{(2\pi)^d} [ip]_J S_{(0)}^{-1\,I\,ij}[ ip]_I,
\ee
which is of the form that we evaluate in the text.

\paragraph{Double insertion trace:} We will also want to compute the trace involving two insertions of the operator  ${\bf S}_{(0)}^{-1} {\bf S}_{(1)}$, which can be evaluated as
\begin{align}
\nonumber
-\frac{i}{4}\Tr({\bf S}_{(0)}^{-1} {\bf S}_{(1)}{\bf S}_{(0)}^{-1} {\bf S}_{(1)}) &= -\frac{i}{4}\int\rd^dp\,\langle p\rvert S_{(0)}^{-1\,I\,ij}[i\hat p]_I\,S_{(1)jk}^{J}(\hat x)[i\hat p]_JS_{(0)}^{-1\,K\,kl}[i\hat p]_K\,S_{(1)li}^{L}(\hat x)[i\hat p]_L\lvert p\rangle\\
&= -\frac{i}{4}\int\frac{\rd^d p}{(2\pi)^d}\frac{\rd^dq}{(2\pi)^d}S_{(0)}^{-1\,I\,ij}[ip]_I \tilde S_{(1)jk}^{J}(p-q)[iq]_JS_{(0)}^{-1\,K\,kl}[iq]_K\,\tilde S_{(1)li}^{L}(q-p)[ ip]_L.
\end{align}
This can be simplified by shifting $q\mapsto q+p$ so that we have
\be
-\frac{i}{4}\Tr\left[({\bf S}_{(0)}^{-1} {\bf S}_{(1)})^2\right]=-\frac{i}{4}\int\frac{\rd^dq}{(2\pi)^d}\tilde S_{(1)jk}^{J}(-q)\tilde S_{(1)li}^{L}(q)\int\frac{\rd^d p}{(2\pi)^d}S_{(0)}^{-1\,I\,ij}[ip]_I [i(q+p)]_JS_{(0)}^{-1\,K\,kl}[i(q+p)]_K[ ip]_L
\ee
The integral over $q$ can now be thought of as a convolution in Fourier space at zero momentum, so transforming back to position space, we obtain:
\be
\label{eq:doubleinsertiontrace}
-\frac{i}{4}\Tr\left[({\bf S}_{(0)}^{-1} {\bf S}_{(1)})^2\right] = -\frac{i}{4}\int\rd^dx S_{(1)li}^{I}(x)\left(\int\frac{\rd^d p}{(2\pi)^d}[\partial+ip]_IS_{(0)}^{-1\,J\,ij}[ip]_JS_{(0)}^{-1\,K\,kl}[\partial+ip]_K[ ip]_L\right)S_{(1)jk}^{L}(x),
\ee
where the derivatives should be understood as acting on $S_{(1)}^J(x)$.

\paragraph{Higher insertions:} This pattern generalizes to higher numbers of insertions. At $n^{\rm th}$ order for an operator at $\ell^{\rm th}$ order in the fields we have
\begin{align}
&\frac{i(-1)^{n-1}}{n}\Tr\left[({\bf S}_{(0)}^{-1} {\bf S}_{(\ell)})^n\right] = \frac{i(-1)^{n-1}}{n}\int\rd^dx \,S_{(\ell)ni}^{I}(x)\\\nonumber
&\times\left(\int\frac{\rd^d p}{(2\pi)^d}[\partial+ip]_IS_{(0)}^{-1\,J\,ij}[ip]_J S_{(0)}^{-1\,K\,kl}[\partial+ip]_KS_{(\ell)jk}^{L}(x)[\partial+ ip]_L\cdots S_{(0)}^{-1\,M\,lm}[\partial+ip]_M S_{(\ell)mn}^{N}(x)[ip]_N\right),
\end{align}
where in the above expression all derivatives should be thought of as acting on {everything} to their right.

\renewcommand{\em}{}
\bibliographystyle{utphys}
\addcontentsline{toc}{section}{References}
\bibliography{galrenorm17}

\end{document}